\documentclass[aps,prl,amsmath,amssymb,superscriptaddress,twocolumn,floatfix]{revtex4-1}
\usepackage{graphicx}
\usepackage{amsmath}
\usepackage{amsfonts}
\usepackage{amssymb}
\usepackage{mathtools}
\usepackage[english]{babel}
\usepackage{xcolor}
\usepackage{hyperref}
\hypersetup{
	colorlinks,
	linkcolor={red!75!black},
	citecolor={blue!75!black},
	urlcolor={blue!75!black}
}

\newcommand{\be}{\begin{equation}}
\newcommand{\ee}{\end{equation}}
\newcommand{\ba}{\begin{array}}
\newcommand{\ea}{\end{array}}

\usepackage{ulem}
\newcommand{\beq}{\begin{eqnarray}}
\newcommand{\eeq}{\end{eqnarray}}

\begin{document}
\title{Gravitational waves from strong first-order phase transitions in an expanding Universe}

\author{Jintao Zou}
\affiliation{~Department of Physics and Chongqing Key Laboratory for Strongly Coupled Physics, Chongqing University, Chongqing 401331, P. R. China}

\author{Zizhuo Zhao}
\affiliation{
~University of Chinese Academy of Sciences (UCAS), Beijing 100049, P.R. China}
\affiliation{
~International Center for Theoretical Physics Asia-Pacific, Beijing/Hangzhou, P.R. China}

\author{Yongtao Jia }
\affiliation{
~University of Chinese Academy of Sciences (UCAS), Beijing 100049, P.R. China}
\affiliation{
~International Center for Theoretical Physics Asia-Pacific, Beijing/Hangzhou, P.R. China}

\author{Ligong Bian\thanks{Corresponding Author.}}
\email{lgbycl@cqu.edu.cn}
\affiliation{~Department of Physics and Chongqing Key Laboratory for Strongly Coupled Physics, Chongqing University, Chongqing 401331, P. R. China}

\begin{abstract}
We investigate the gravitational waves generated by first-order phase transitions (PTs) in an expanding Universe. Using a scalar field-fluid coupled model, we perform three-dimensional lattice numerical simulations of slow first-order PTs with strong strength. We observe that the contribution from sound waves to the gravitational wave (GW) signal significantly dominates over that from bubble collisions. The sound wave spectra exhibit two peaks located at the mean bubble separation and the sound-shell width. Rotational modes are generated in the fluid during the bubble collisions, and their magnitude grows with the PT strength ($\alpha$). The effective lifetime of the sound-wave source increases for slower PTs, decreasing for faster PT with larger $\beta/H$ and rising with smaller $\alpha$. While the simulated amplitude of normalized gravitational waves is of order $\mathcal{O}(10^{-2})$, in alignment with Minkowski-spacetime results, the resulting GW spectra differ markedly from those scenarios.

\end{abstract}	

\maketitle

\noindent{\it \bfseries Introduction.}
Cosmological first-order phase transition (PT) is of significant importance and has implications in particle physics and cosmology, such as: serving as the cosmic origin of the primordial magnetic field~\cite{Hogan:1983zz,Di:2020kbw,Enqvist:1993np,Yang:2021uid,Vachaspati:1991nm,Di:2024gsl,Di:2025ncl,Baym:1995fk,Grasso:1997nx,Quashnock:1988vs}, producing the primordial black hole~\cite{Liu:2021svg,Franciolini:2025ztf,Carr:2026hot,Lewicki:2023ioy,Gouttenoire:2023naa,Jedamzik:1999am}, providing the thermal departure condition for explanation of the matter-antimatter asymmetry of the universe~\cite{Morrissey:2012db, Cohen:2012zza}, and the production of gravitational waves (GWs) that is the primary target of space-based interferometers~\cite{Mazumdar:2018dfl, Caprini:2015zlo,Caprini:2019egz}, such as: Laser Interferometer Space Antenna (LISA)~\cite{LISA:2017pwj}, Taiji~\cite{Caprini:2019egz}, and TianQin~\cite{TianQin:2015yph}.

The determination of the GW spectra from PTs necessitates numerical simulations~\cite{Athron:2023xlk}, since the nonlinear effects are significant during such a violent process. When the PT process is fast enough, the Hubble expansion effect has been assumed to be negligible. In this limit, great progress have been made in the determination of the GW spectra from vacuum bubble collisions~\cite{Cutting:2018tjt,Cutting:2020nla,Zhong:2021hgo} and from sound waves~\cite{Hindmarsh:2013xza,Hindmarsh:2015qta,Hindmarsh:2017gnf,Cutting:2019zws,Giblin:2014qia,Cutting_2022,Giombi:2023jqq,Giombi:2025tkv,Wang:2026jnh}, complemented by the sound shell model~\cite{Hindmarsh:2018ssm,Hindmarsh:2019ssm}. These results suggest that the sound waves dominantly produce the GWs from PTs in the thermal plasma. Beyond the spectral shape, the temporal evolution of the sound wave source has also attracted considerable attention, the decay of the acoustic turbulence generated by the sound waves has been systematically studied in two~\cite{Dahl:2021wyk} and three dimensions~\cite{Dahl:2024eup,Auclair:2022turb}.

The Hubble expansion effect, however, would not be negligible, since the equation of state of the Universe is changing during the PT reheating process, as suggested by recent studies~\cite{Yamada:2025hfs,Yamada:2025cfr,Zou:2026wzi,Zou:2025sow,Zhao:2022cnn,Guo:2024kfk}. Therefore, the real-time history evolution with the expansion of the Universe is crucial. A first attempt was made in Ref.~\cite{Kurki-Suonio:1996gkq}, where the PT was simulated in a spherically symmetric expanding background based on the field-fluid coupled equations developed in the flat Universe~\cite{Kurki-Suonio:1995yaf}, with the bubble collision effects captured through the reflective boundary conditions at the (spherical) outer edge; the reheating phenomenon and the influence of the potential and the friction parameter $\eta$ on the bubble wall velocity were observed. More recently, the finite lifetime of the sound wave source has been pointed out~\cite{Ellis:2020awk,Guo:2020grp}, together with the signatures of the speed of sound on the GW power spectrum~\cite{Giombi:2024kju}.

In this Letter, we investigate the effect of the Hubble expansion through numerical simulations of the thermal first-order PT process in an expanding Universe. Unlike previous studies, we adopt a more realistic scenario in which the PT strength and the equation of state evolve dynamically during the PT process. We perform three-dimensional lattice numerical simulations for various values of the PT strength $\alpha$ and the inverse transition $\beta/H$. In all our simulations, the sound wave contribution dominates over that from bubble collisions. For the GWs sourced by sound waves, we observe a double peak feature. Furthermore, the ultraviolet (UV) tail of the GW spectrum for deflagrations is steeper than that for detonations. We work in natural units with $c = 8\pi G = 1$.

\noindent{\it \bfseries Simulation setup.}
The field-fluid model is taken to study the PT dynamics that occurred in the early universe, which couples an order parameter field to the fluid through a friction term. This system has three fundamental variables: the local temperature $T$, the 4-velocity of fluid $U^\mu$ and the scalar field $\phi $. The energy-momentum tensor of system is $T^{\mu \nu } = \partial^{\mu} \phi \partial^{\nu} \phi - \frac{1}{2} g^{\mu \nu} (\partial \phi)^2 + (\epsilon + p ) U^\mu U^\nu +  g^{\mu \nu} p $, where the metric convention is $\left(-,+,+,+\right)$, and can be divided into two parts with the friction term $ \delta^{\nu} = \eta U^{\mu} \partial_{\mu} \phi \partial^{\nu} \phi $. Generalizing the Minkowski spacetime case of Refs.~\cite{Hindmarsh:2015qta,Hindmarsh:2017gnf}, the conservation of the energy-momentum tensor yields the equations of motion in the FLRW spacetime~\cite{Guo:2020grp}
\begin{equation}
\begin{aligned}
& -\ddot{\phi} + \frac{1}{a^2}\nabla^2 \phi - \frac{\partial V}{\partial \phi } - 3H\dot{\phi} = \eta \gamma \left(\dot{\phi} + \frac{1}{a}v^{i}\partial_{i} \phi \right), \\
&\dot{E}+3H(E+\gamma p)+\frac{1}{a}\partial_i \left(E v^{i}\right) + p \left[ \dot{\gamma}+\frac{1}{a}\partial_i \left(\gamma v^{i}\right) \right], \\
\end{aligned}
\end{equation}
\begin{equation}
\begin{aligned}
&\quad - \frac{\partial V}{\partial \phi} \gamma \left(\dot{\phi} + \frac{1}{a} v^{i}\partial_i \phi\right) = \eta \gamma^2 \left(\dot{\phi} + \frac{1}{a} v^{i}\partial_i \phi\right)^2\;, \\
&\dot{Z_i} + 5HZ_i + \frac{1}{a}\partial_j\left( Z_i v^j\right)+ \frac{1}{a^2} \partial_i p + \frac{1}{a^2}\frac{\partial V}{\partial \phi }\partial_i \phi \\
&\qquad\qquad\qquad
= -\frac{1}{a^2}\eta \gamma\left(\dot{\phi}  + \frac{1}{a} v^{j}\partial_j \phi\right) \partial_i \phi \;,
\end{aligned}
\end{equation}
where the fluid energy density is $E=\epsilon\gamma$, with $\gamma=1/\sqrt{1-\vec{v}^2}$ being the Lorentz factor, and the fluid momentum density is $Z_i=\gamma(\epsilon+p)U_i$, with fluid 4-velocity $U^{\mu}=\gamma(1,\vec{v}/a)$. In this paper, $H$ denotes the physical Hubble parameter and $t$ the physical time with $H^2=(\frac{\dot{a}}{a})^2=\frac{\rho}{3}$, while $\mathcal{H}$ and $\tau$ denote the comoving Hubble parameter and the comoving time. We take friction parameter $\eta=\tilde{\eta}\frac{\phi^2}{T}$, a form better motivated by the underlying microphysics~\cite{Moore:1995si,Huber:2013kj,Hindmarsh:2017gnf}. For a given wall speed and transition strength, this choice only mildly affects the fluid velocity around the bubbles, which is determined purely by hydrodynamics~\cite{Hindmarsh:2017gnf,Espinosa:2010hh,Kurki-Suonio:1995yaf,Steinhardt:1981ct}. 

The simulated vacuum energy density is given by $V_{\rm sim}(\phi,T)=\varepsilon_0\,3a\,T^4\,\bar V(\bar\phi,\bar T)$, where $\bar\phi=\phi/\phi_b$, $\bar T=T_0/T_c$, $T_c$ is the critical temperature, $\phi_b$ is the true vacuum field value, $a = \left(\pi^2/90\right)g_{\ast}$ and $ g_{\ast}=106.25$ is the Standard Model effective number of degrees of freedom. The dimensionless effective potential is motivated by the new physics beyond Standard Model that introduce dimension six operator to trigger a first-order PT~\cite{Chung:2012vg}, which takes the form 
\begin{equation}
    \bar V(\bar\phi,\bar T)=\frac{-m^2+\lambda_2\bar T^2}{2}\bar\phi^2+\lambda_4\bar\phi^4+\lambda_6\bar\phi^6,
\label{eq:potential}
\end{equation}
with parameters $m^2=0.611$, $\lambda_2=2.25$, $\lambda_4=-4.996$, $\lambda_6=4.307$. Here, the parameter $\varepsilon_0$ is fixed by requiring the PT strength $\alpha\sim0.1$ at $T_0=0.8T_c$. In our simulations, $T_c=0.0064$, two sets of field values are adopted: $\phi_b=0.156\,T_c$ and $\phi_b=0.625\,T_c$, whose simulation results correspond to the detonation and deflagration regimes, respectively. The squared sound speed is defined by $c_s^2 = dp/d\epsilon$, evaluated along the equilibrium equation‑of‑state. Starting from
\begin{equation}
    \epsilon = 3 a T^4 + V(\phi,T) - T\frac{\partial V_{sim}}{\partial T} , \quad p = a T^4 - V_{sim}(\phi,T).
\end{equation}
The total entropy density reads $s_{\rm tot}(T) = dp/dT = 4aT^3 - V_{sim}'(T)$, Differentiating $\epsilon(T)$ with respect to $T$ gives the constant-volume heat capacity $\frac{d\epsilon}{dT} = 12aT^3 - TV_{sim}''(T)$, Substituting into $c_s^2$, one has
\begin{equation}
c_s^2 = \frac{4aT^3 - V_{sim}'(T)}{12aT^3 - TV_{sim}''(T)}.
\label{eq:c_s}
\end{equation}

Through the Einstein equation, we can get equation of motion for tensor perturbation $h_{ij}$
\begin{align}
	\ddot{h}_{i j}-\frac{1}{a^2}\nabla^2 h_{i j}+3H\dot{h}_{ij}=\frac{16 \pi G}{a^2} \tau_{i j}^{\mathrm{TT}},
\end{align}
where $\tau_{i j}^{\mathrm{TT}}$ is the transverse traceless part of the stress-energy tensor from scalar field and fluid. Each of them can be obtained from $\tau_{ij}^{TT} = \tau_{ij}^{\phi} + \tau_{ij}^\mathrm{f}$, with the field part $\tau_{ij}^{\phi} = \partial_i \phi \partial_j \phi$ and the fluid part $ \tau_{ij}^\mathrm{f} = \gamma^2\left(\epsilon + p \right) v_i v_j/a^2$. With which, we can get fractional energy density in GWs from scalar field and fluid parts
\begin{align}\label{GW_energy}
    \Omega_{gw}^{f,\phi}(k)&=\frac{1}{\rho}\frac{d\rho_{gw}^{f,\phi}}{d\ln k} 
    = \frac{1}{24\pi^2 V}\frac{k^3}{H^2}\sum_{i,j}|\dot{h}_{ij}(k,t)|^2,
\end{align}
where $V$ is the total volume. $k$ is the magnitude of wave vector and $h_{ij}(k,t)$ is the Fourier transformation of $h_{ij}(r,t)$. The dimensionless GWs spectrum can be estimated as \cite{Hindmarsh:2015qta,Hindmarsh:2017gnf,Hindmarsh:2020hop}
\begin{align}\label{eq:Omega_parameter}
    \Omega_{gw}^{f}=3(1+\omega)^2 \overline{U}_{f}^4 (\mathcal{H}_n R^*)\Upsilon\tilde{\Omega}_{gw}^{f},
\end{align}
where the effective lifetime $\tau_v$ of the sound-wave source is defined through $\mathcal{H}\tau_v\equiv\Upsilon$. Since the equation of state $\omega=p/\epsilon$ varies in our simulations, we use the general $\omega$ form~\cite{Xiao:2024rsj}
\begin{equation}
    \Upsilon = \frac{2\left[1 - y^{3(\omega-1)/2}\right]}{3(1-\omega)},
    \label{eq:Htau_sw}
\end{equation}
where $y=a(t)/a(t_s)$ is the ratio of the scale factor at a later time $t$ to that at $t_s$ when GW production from sound waves starts for $\omega=1/3$ this reduces to $\mathcal{H}\tau_v=1-1/y=1-\frac{1}{\sqrt{1+2\mathcal{H}\tau_{sh}}}$~\cite{Hindmarsh:2020hop}. The factor saturates at $2/[3(1-\omega)]$, which is less than unity for $\omega<1/3$. Consequently, even when $\mathcal{H}\tau_{sh}=R^*\mathcal{H}/\overline{U}_f$ exceeds a Hubble time, $\Upsilon$ always remains below 1. This indicates that the effective source lifetime is always shorter than a Hubble time.

Here, the root mean square fluid velocity $ \overline{U}_\mathrm{f}$ is defined by $ \left( \overline{\epsilon} + \overline{p} \right)\overline{U}_{f}^{2} = \frac{1}{\mathcal{V}}\int_{\mathcal{V}} d^3 x \tau_{ii}^\mathrm{f}$ in which $\overline{\epsilon}$ and $ \overline{p} $ correspond to spatially averaged energy density and pressure and $\mathcal{V} $ is the averaging volume. And for scalar field is the same $ \left( \overline{\epsilon} + \overline{p} \right)\overline{U}_{\phi}^{2} = \frac{1}{\mathcal{V}}\int_{\mathcal{V}} d^3 x \tau_{ii}^{\phi}$. Noting that the above definition uses covariant vector $v_i$ but in equation of motion, we use contravariant vector $v^i$. So using $v_i=g_{ij}v^j=a^2v^i$, we have
\begin{equation}\label{eq:UfUphi}
\begin{aligned}
\left( \overline{\epsilon} + \overline{p} \right)\overline{U}_{f}^{2} &= \frac{1}{\mathcal{V}}\int_{\mathcal{V}} d^3 x \gamma^2(\epsilon+p)a^2\vec{v}^2\;, \\
\left( \overline{\epsilon} + \overline{p} \right)\overline{U}_{\phi}^{2} &= \frac{1}{\mathcal{V}}\int_{\mathcal{V}} d^3 x (\nabla \phi)^2\;.
\end{aligned}
\end{equation}
The mean squre veolicity of the fluid connect with the PT strength $\alpha$ as~\cite{Hindmarsh:2017gnf,Ellis:2020awk,Cai:2018teh}
\begin{equation}\label{eq:kappa}
    (1+w)\overline{U}_{\mathrm{f}}^2 = \frac{\kappa\alpha}{1+\alpha},
\end{equation}
where the parameter $\kappa$ is defined as the efficiency with which vacuum energy is converted into fluid kinetic energy that can be extracted from the simulation results, and the PT strength is defined as $\alpha = \frac{\Delta\theta(T)}{3a_{eff} T^4}$, where $\Delta\theta(T)=\frac{1}{4}\frac{d}{dT}\Delta V_{sim}-\Delta V_{sim}$, with $\Delta V_{sim}$ being the potential difference between the true and false vacua. 

All simulations are performed in the expanding Universe on a three-dimensional lattice of size $L^3=(512dx)^3$ corresponding to a physical volume of $LH=3$. The numerical calculations are performed using a code developed based on {\it pystella}~\cite{Weiner:2021zj}. The values of the two key PT parameters, the PT strength $\alpha$ and the inverse transition rate $\beta/H$, are summarized in Tab.~\ref{tab:wide_table}. The initial temperature $T_0$, the initial bubble radius $R_0$, and the wall thickness $l_w$ for each parameter set are also listed in Tab.~\ref{tab:wide_table}. The $R_0$ and $l_w$ are obtained using the {\it FindBounce}~\cite{Guada:2020xnz}. The PT starts with the nucleation of true-vacuum bubbles following the exponential nucleation rate $p(t) = p_f e^{\beta \left( t - t_f\right)}$, with $p_f = 400\beta^4$ and $t_f = 0$. Since both $\phi$ and $T$ evolve during the PT, we perform simulations with varing PT strength, $\eta$ is also a dynamical quantity that varies in space and time. In our simulations, we take $\tilde{\eta}=500\sim T_0/\phi_b^2$, see the {\it The evolution of the PT parameters} section of the {\it Supplemental Material}, and all PT‑strength parameters $\alpha$ in the main text refer to the initial value. The mean bubble separation is obtained as the bubble radius $R^*=(L^3/N_b)^{1/3}$, where $N_b$ is the number of bubbles generated during the PT. For fast bubble wall velocities, we obtain $v_w$ via $\gamma = R^*/2R_0 = 1/\sqrt{1-v_w^2}$. For slow bubble wall velocities, we calculate $v_w$ using $v_w = \sqrt{\rho_K/\rho_G}$~\cite{Cutting:2022zgd}, where $\rho_K$ is the kinetic energy density and $\rho_G$ is the gradient energy density.
\begin{table}[!htp]
    \centering
    \renewcommand{\arraystretch}{1.2}
    \setlength{\tabcolsep}{2pt}  
    \caption{Bubble parameters for various values of $\beta/H$ and $\alpha_0$. The left columns present the type of PT (Deto. for detonation, Defl. for deflagration), the PT strength $\alpha$, and the number of generated bubbles $N_b$ for $\beta/H$ = 20, 25, 30. The right columns list the parameter $\varepsilon_0$, the initial temperature $T_0/T_c$, initial bubble radius $R_0$, wall thickness $l_w$, and bubble wall velocity $v_w$.}
    \begin{tabular}{ccccc|ccccc}
        \hline
        Type   &$\alpha_0 \backslash \beta/H$   &       20       &       25   &       30       & $\varepsilon_0$ & $T_0/T_c$  &$R_0/d x$& $l_w/dx$ &$v_w$ \\
        \hline
        Deto.  &$  0.15  $                     & 322 & 336 & 219 & 0.683 & 0.084 & 22  & 14 & 0.81   \\
               &$  0.38  $                     & 525 & 649 & 690 & 0.556 & 0.191 & 13  & 16 & 0.91   \\
               &$  0.60  $                     & 608 & 791 & 926 & 0.508 & 0.273 & 9   & 18 & 0.95   \\
        \hline
        Defl.  &$  0.15  $                     & 314 & 339 & 356 & 0.683 & 0.084 & 22  & 14 & 0.27   \\
        \hline
    \end{tabular}
    \label{tab:wide_table}
\end{table}

\begin{figure}[!htp]
    \centering
    \includegraphics[width=0.45\textwidth,clip]{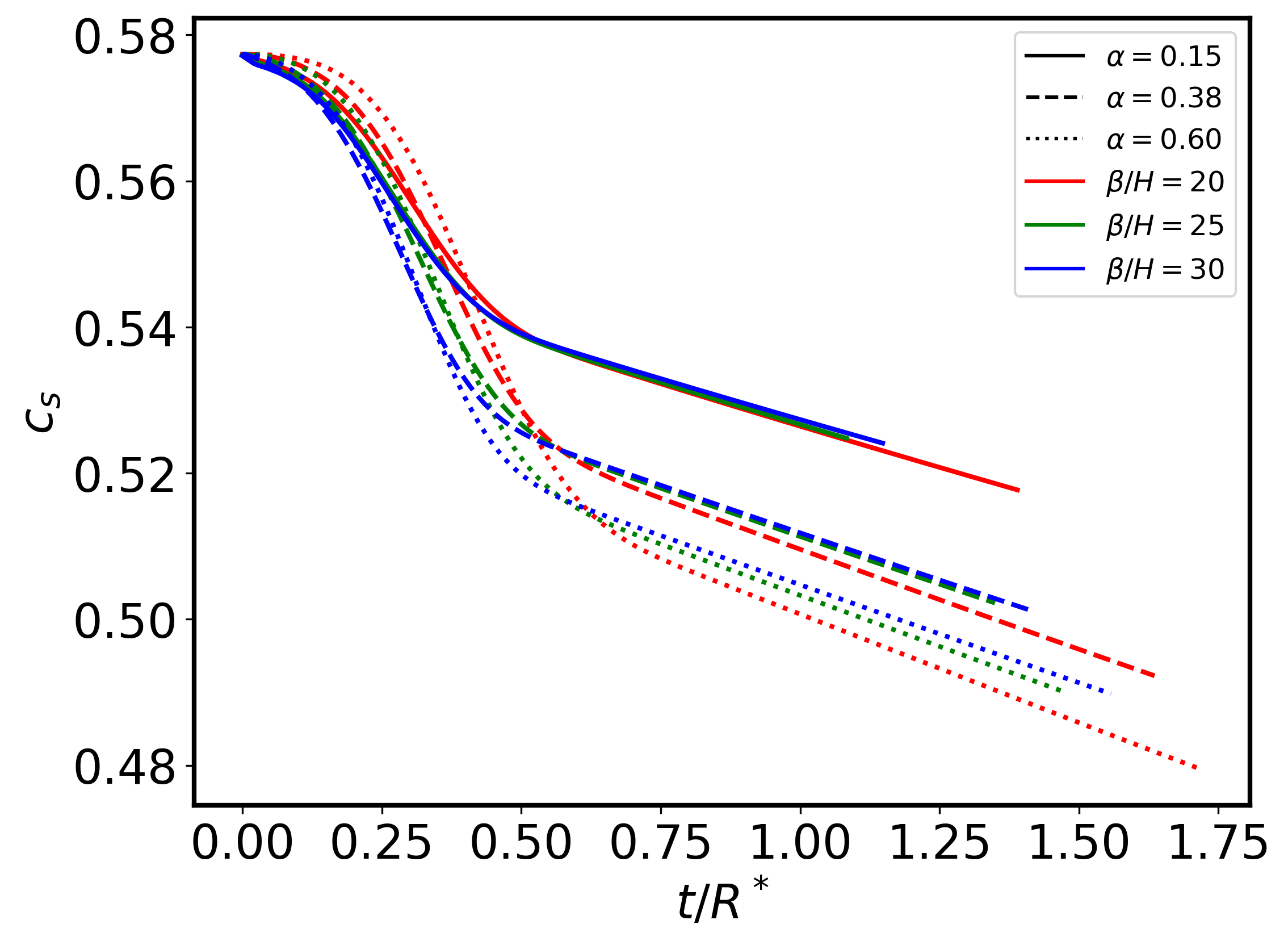}
    \includegraphics[width=0.45\textwidth,clip]{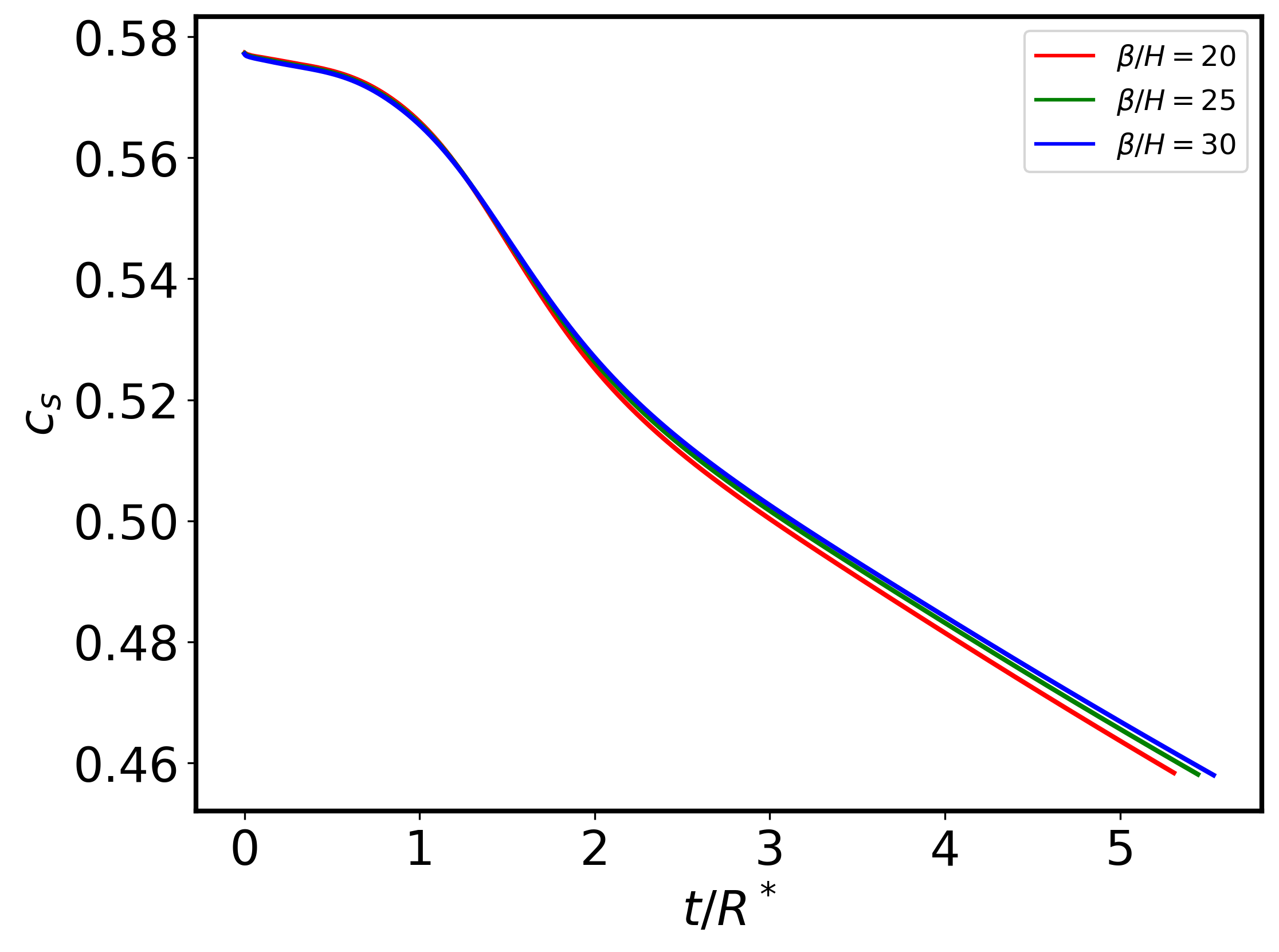}
    \caption{The evolution of the sound speed $c_s$ in the expanding Universe for detonations (top) and deflagrations (bottom). Line styles distinguish $\alpha_0 = 0.15$ (solid), $0.38$ (dashed) and $0.60$ (dotted), while red, green and blue correspond to $\beta/H = 20$, $25$ and $30$.}
\label{fig:c_s}
\end{figure}

\begin{figure}[!htp]
    \centering
    \includegraphics[width=0.45\textwidth,clip]{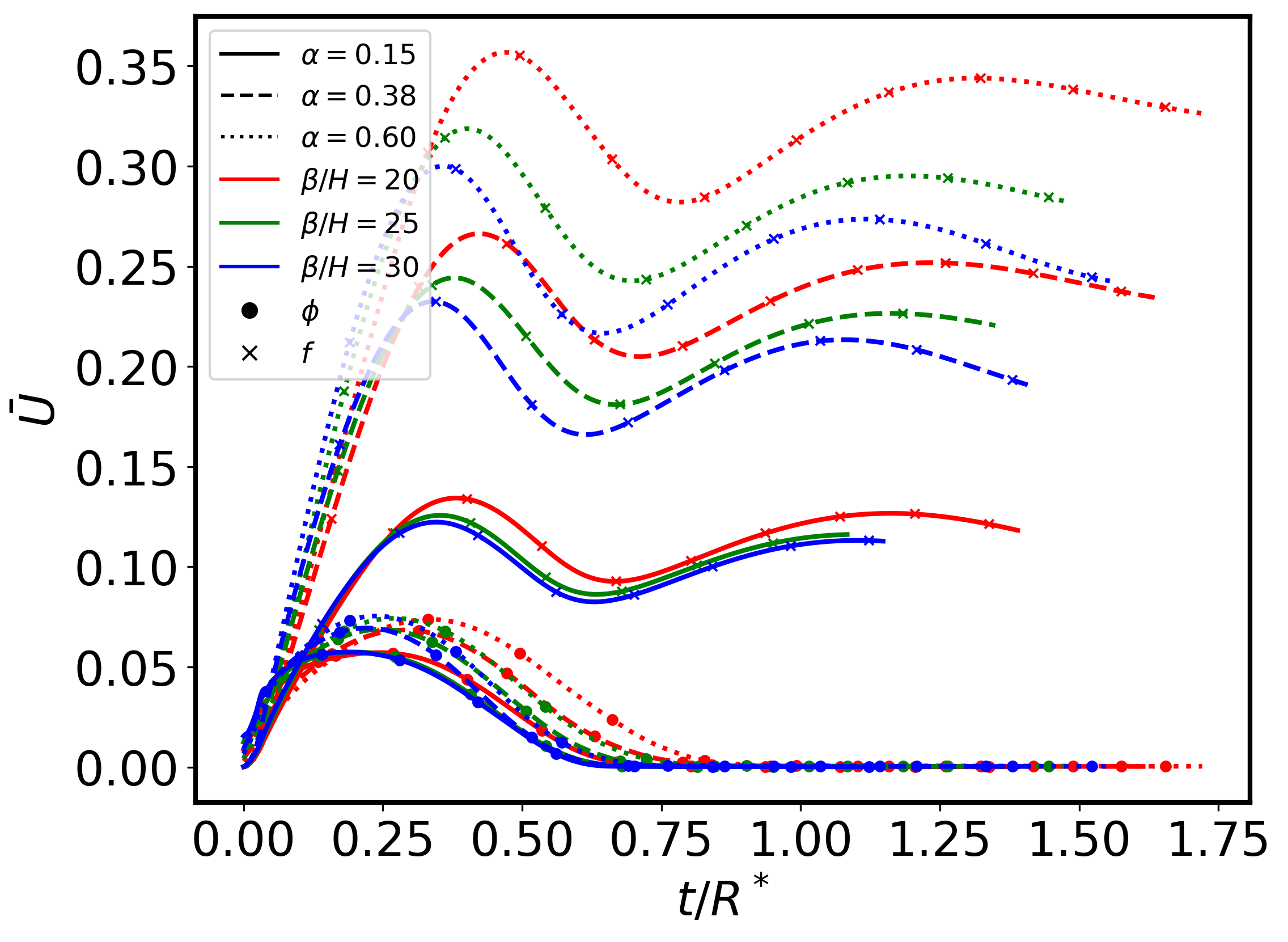}
    \includegraphics[width=0.45\textwidth,clip]{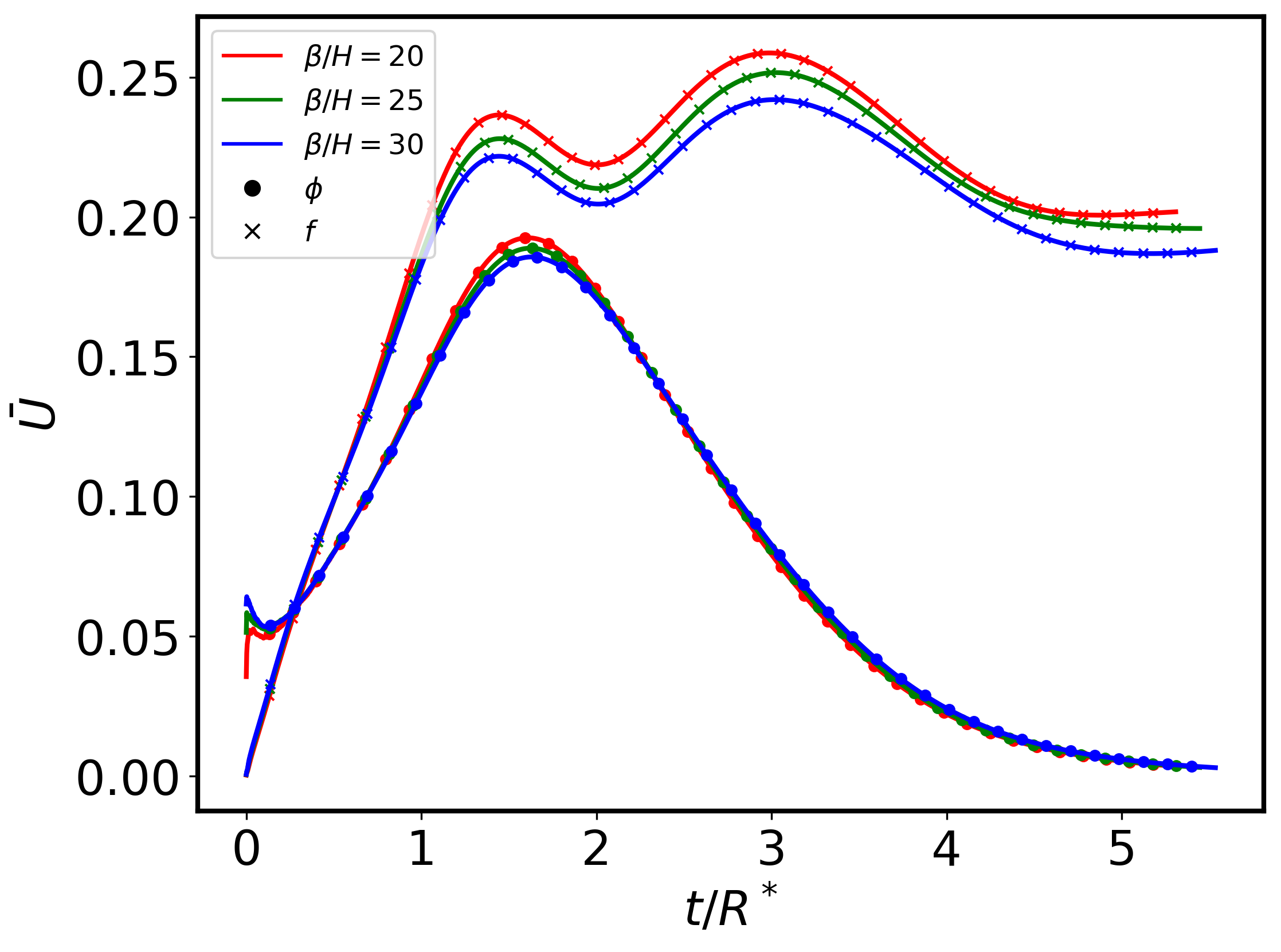}
    \caption{Mean square velocities of the field ($\overline{U}_{\phi}$, dots) and fluid ($\overline{U}_f$, crosses) in the expanding Universe, for detonation scenarios (top panel) and deflagration scenarios (bottom panel). Line styles distinguish $\alpha_0 = 0.15$ (solid), $0.38$ (dashed) and $0.60$ (dotted), while red, green and blue correspond to $\beta/H = 20$, $25$ and $30$.}
\label{fig:U}
\end{figure}

\noindent{\it \bfseries Simulation Results.}
For the GW spectral shape, $c_s$ fixes the width of the sound shell, $\Delta R^*=|v_w-c_s|/v_wR^*$~\cite{Hindmarsh:2019ssm}, which determines the positions of the peaks of the sound wave spectrum discussed below. Therefore, we first evaluate the sound speed using Eq.~(\ref{eq:c_s}) and show the evolution of $c_s$ under different scenarios in Fig.~\ref{fig:c_s}. For the parameters we adopt, $c_s$ is always smaller than $1/\sqrt{3}$, and a larger $\alpha_0$ corresponds to a smaller $c_s$. In our calculations, the sound speed mainly affects two aspects. First, in fluid evolution, the sound speed $c_s^2=dp/d\epsilon$ sets the speed at which pressure perturbations propagate relative to the plasma, thereby controlling how the pressure-gradient force drives the fluid and converts the released vacuum energy into fluid motion. Second, since $c_s$ is the sound speed in the plasma, walls with $v_w<c_s$ propagate as deflagrations, whereas those with $v_w>v_J$ propagate as detonations, where $v_J>c_s$ is the Chapman-Jouguet velocity~\cite{Hindmarsh:2020hop}. With $c_s\simeq0.46$-$0.58$, our runs with $v_w\simeq0.81$-$0.95$ ($0.27$) therefore belong to the detonation (deflagration) classifications~\cite{Guo:2020grp,Espinosa:2010hh}.

Then we present the mean square velocities of the field and fluid, $\overline{U}_{\phi}$ and $\overline{U}_{f}$ since they characterize the magnitude of the GW spectra. As shown in Fig.~\ref{fig:U}, both $\overline{U}_f$ and $\overline{U}_{\phi}$ initially increase as bubbles expand. Energy transfer from the field to the fluid then causes $\overline{U}_{\phi}$ to decrease while $\overline{U}_f$ increases, eventually reaching near-equilibrium. $\overline{U}_f$ exhibits a dip around $t/R^*\simeq 0.7$ in the detonation case and around $t/R^*\simeq 2$ in the deflagration case, due to the variation of $\alpha$; its evolution is shown in the {\it Supplemental Material}. We observe that both $\overline{U}_f$ and $\overline{U}_{\phi}$ grow with the PT strength $\alpha$, since a stronger transition transfers more energy into the kinetic motion of the field and fluid~\cite{Hindmarsh:2017gnf}, leading to larger peaks of both $\overline{U}_\phi$ and $\overline{U}_f$. For a fixed $\alpha$, increasing $\beta/H$ shifts the peaks of both $\overline{U}_{\phi}$ and $\overline{U}_f$ to a smaller $t/R^*$, but reduces their peak values. A larger $\beta/H$ corresponds to a faster PT, which generates more bubbles more rapidly (Tab.~\ref{tab:wide_table}) and fills the Universe earlier, so the bubble walls collide and annihilate at an earlier time. Since the kinetic energy of the field resides predominantly in the bubble walls, $\overline{U}_\phi$ and $\overline{U}_f$ peak when the wall density is largest, which is at the moment of most violent collisions. Therefore, for a larger $\beta/H$, this moment occurs at a smaller $t/R^*$. The kinetic energy of the fluid is transferred from the released potential energy through the work done by the moving walls on the plasma. A shorter duration before bubble collision means the walls have less time to release potential energy before merging, leading to smaller peaks of both $\overline{U}_\phi$ and $\overline{U}_f$.
\begin{figure}[!tp]
    \includegraphics[width=0.45\textwidth,clip]{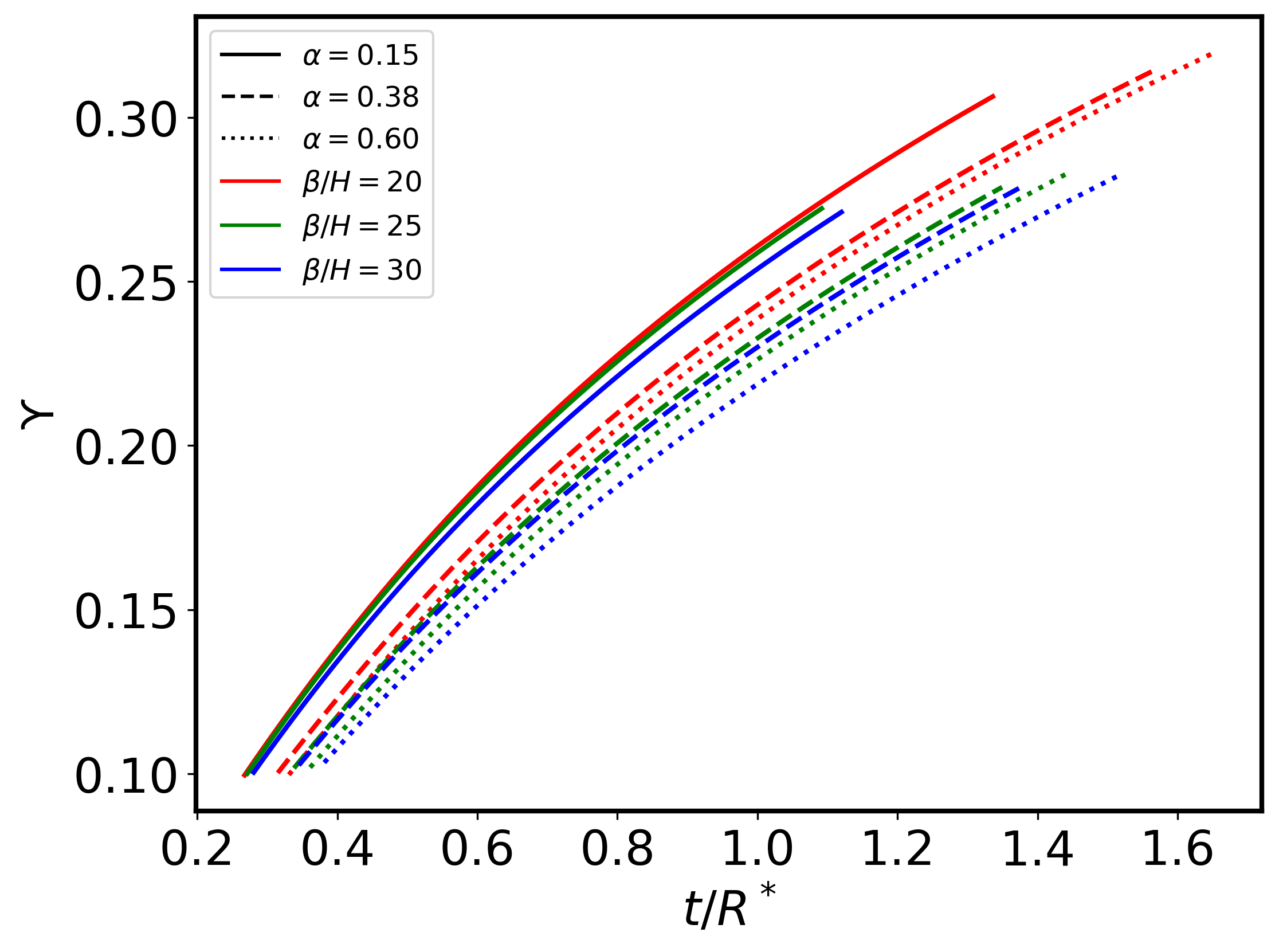}
    \includegraphics[width=0.45\textwidth,clip]{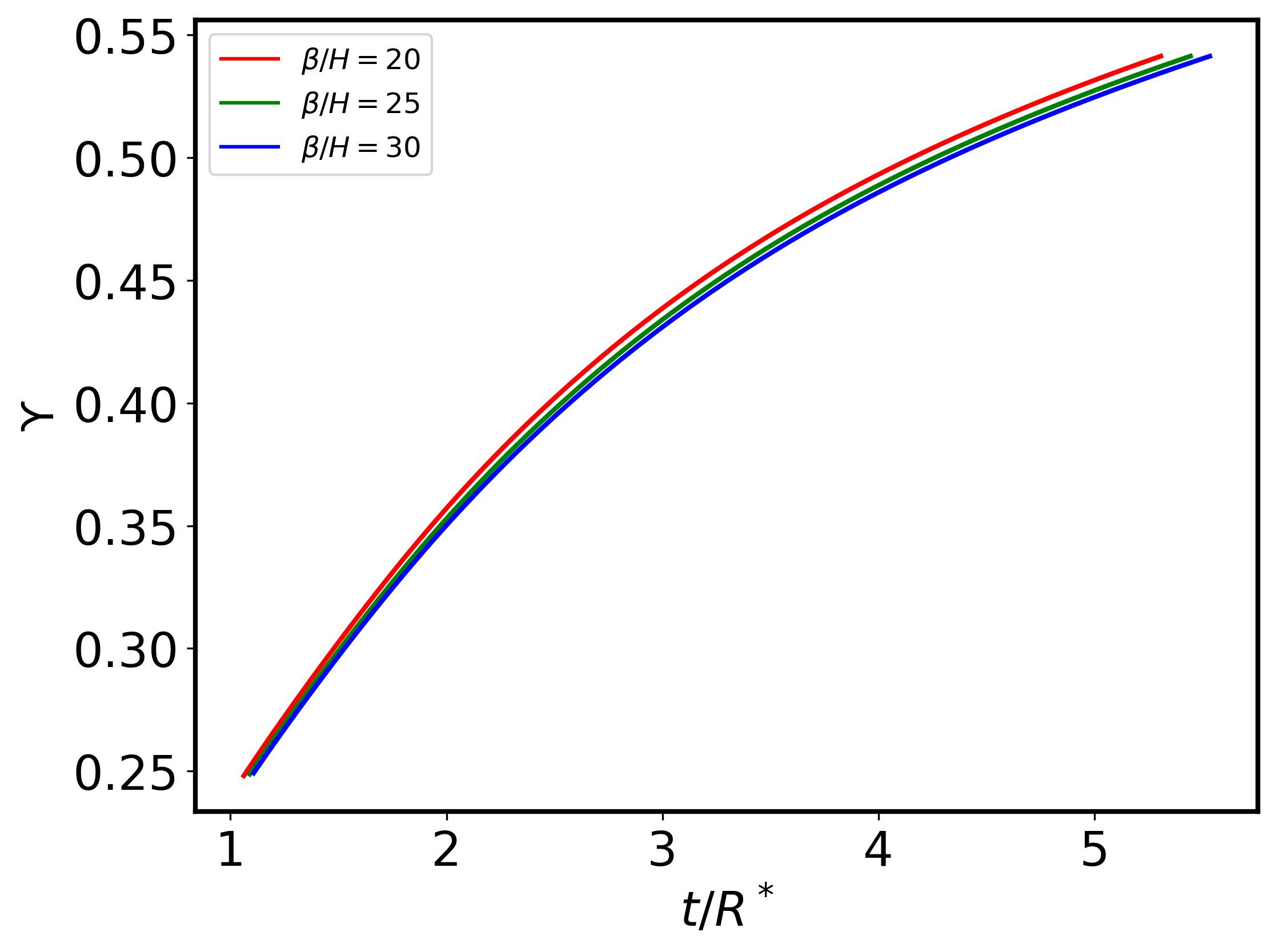}
   	\caption{The evolution of growth factor $\Upsilon$ in the expanding Universe for detonation scenarios (top) and deflagration scenarios (bottom). Line styles distinguish $\alpha_0 = 0.15$ (solid), $0.38$ (dashed) and $0.60$ (dotted), while red, green and blue correspond to $\beta/H = 20$, $25$ and $30$.}
    \label{fig:Htau}
\end{figure}
\begin{figure}[!tp]
    \centering
    \includegraphics[width=0.45\textwidth,clip]{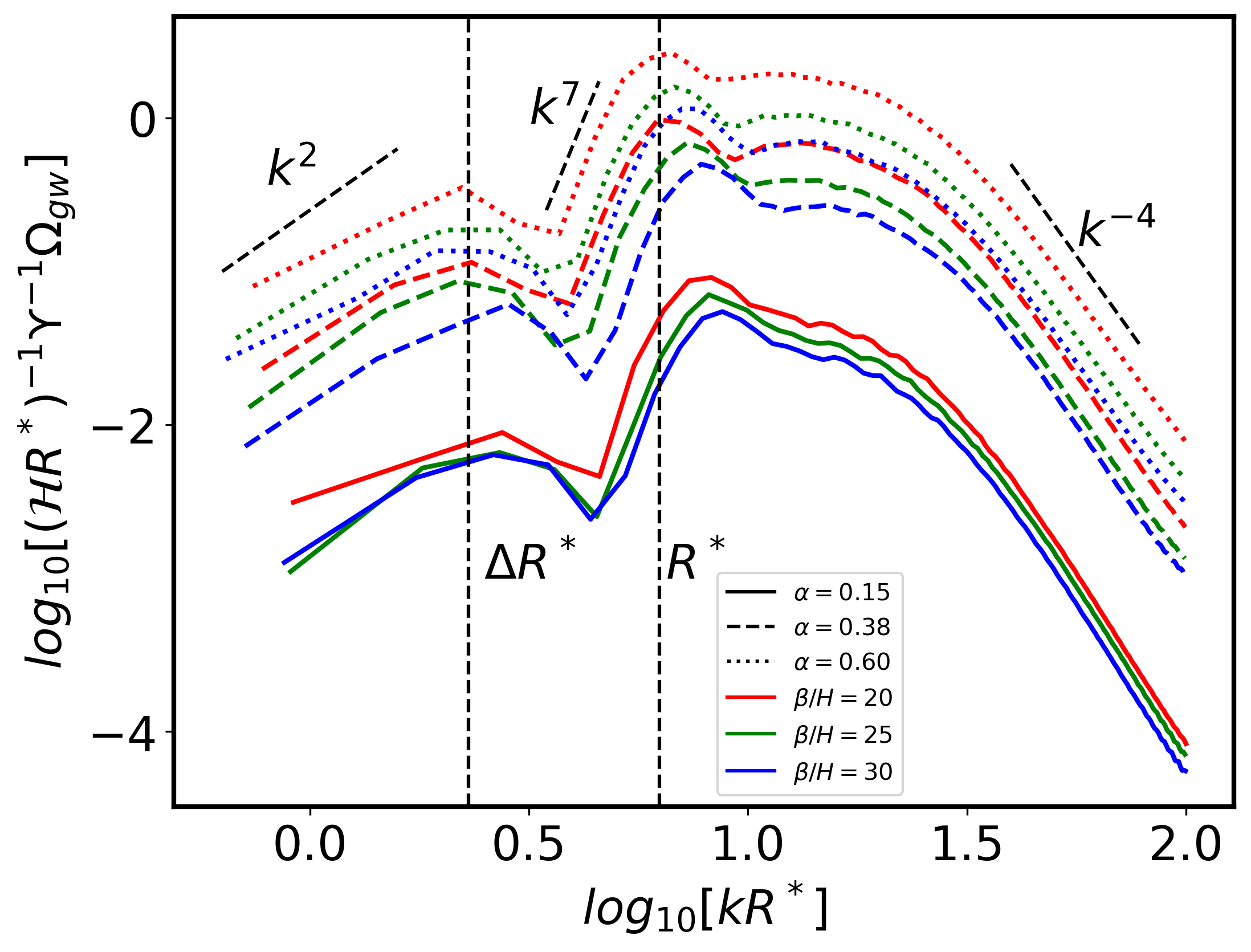} \\
    \includegraphics[width=0.45\textwidth,clip]{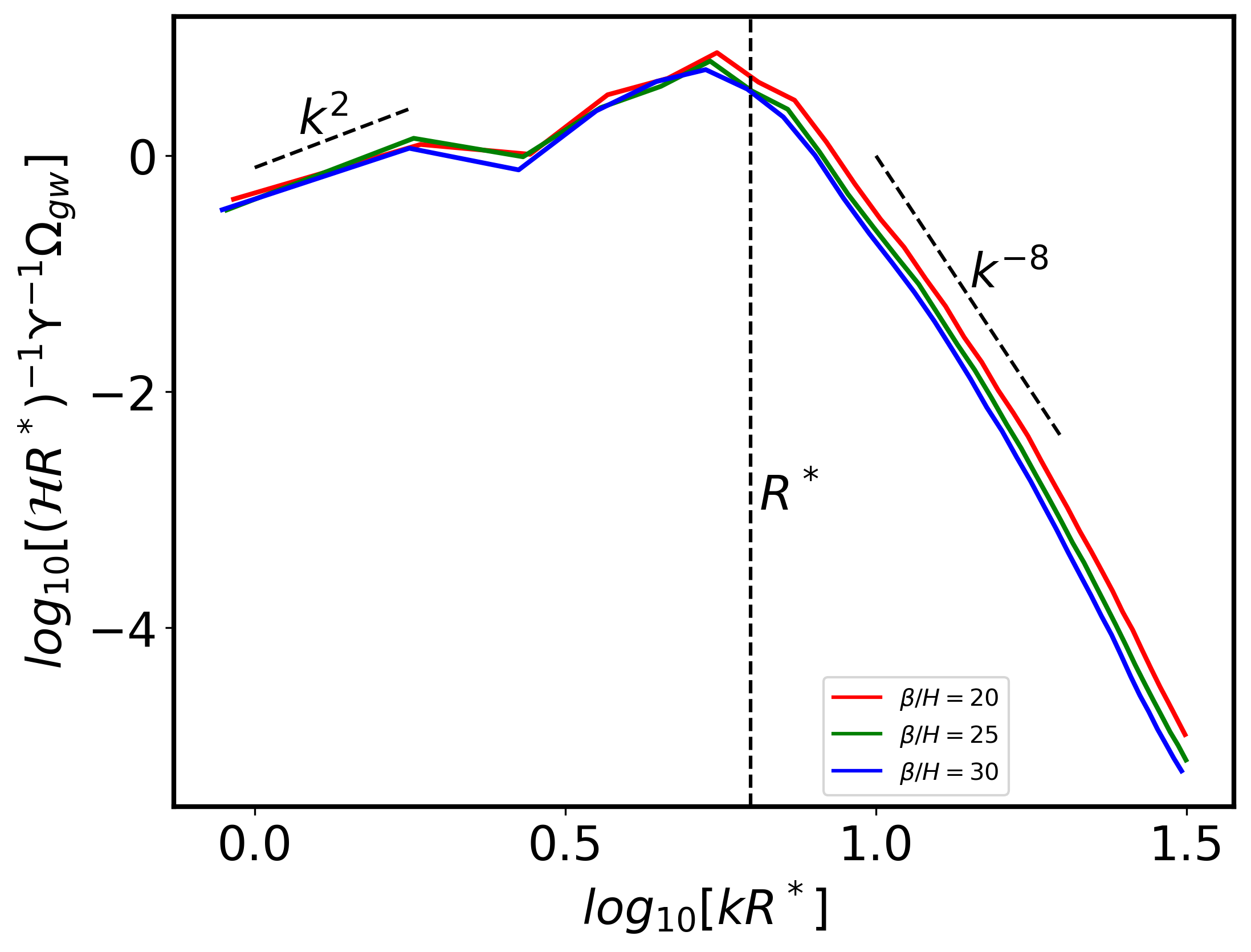}
   	\caption{The normalized sound wave (fluid) GW spectra after the spectral shape has stabilized for detonations (upper) and deflagrations (lower). In the upper panel, line styles distinguish $\alpha_0 = 0.15$ (solid), $0.38$ (dashed) and $0.60$ (dotted), and red, green and blue correspond to $\beta/H = 20$, $25$ and $30$.}
    \label{fig:all_gw}
\end{figure}

To study the evolution of GWs from sound wave sources in an expanding universe, we evaluate the growth factor $\Upsilon$ introduced in Refs.~\cite{Guo:2024kfk,Guo:2020grp}, and its evolution is presented in Fig.~\ref{fig:Htau}. We find that $\Upsilon$ decreases with increasing $\alpha$, which is because a larger $\alpha$ implies greater vacuum energy in our simulations, causing the scale factor $a$ to grow more rapidly. Furthermore, $\Upsilon$ decreases with increasing $\beta/H$, primarily due to the influence of $\omega$, the evolution plot of $\omega$ is provided in the {\it Supplemental Material}.

Finally, we present in Fig.~\ref{fig:all_gw} the sound wave (fluid) GW spectra, normalized by $(\mathcal{H}_n R^*)^{-1}\Upsilon^{-1}$, for different values of $\alpha$ and $\beta/H$ after the spectral shape has stabilized. In the infrared, all spectra rise as $k^2$. Causality arguments typically require the infrared slope to follow a $k^3$ power law, but this holds only for modes that are still super-horizon when the source disappears or stops oscillating~\cite{Cai:2019cdl,Caprini:2018mtu}. In our simulations, once the bubble walls collide and transfer their energy to the fluid, the sound waves continue to oscillate and act as a source of GWs up to the Hubble time. As a result, $\Omega_{GW}$ in the infrared follows $k^2$ rather than $k^3$. Because the bubble collisions and the fluid motion are coupled in energy, the collision (field) contribution to the GWs also departs from the $k^3$ slope. In principle, $k^3$ should be recovered for modes at still lower wavenumbers that remain super-horizon throughout the lifetime of the source; however, resolving such low $k$ lies beyond the dynamic range of the present simulations and cannot be observed directly. For detonation scenarios, the sound wave spectra exhibit a double-peak structure, with a steep $k^7$ segment between the two peaks, followed by a $k^{-4}$ decay in the ultraviolet. The peak amplitude of the sound wave spectrum increases with $\alpha$ and decreases with increasing $\beta/H$. Quantitatively, the normalized peak amplitude varies by more than an order of magnitude across the detonation runs, from $\mathcal{O}(10^{-2})$ ($\alpha_0=0.15$) to $\mathcal{O}(1)$ ($\alpha_0=0.60$), and lies with $\mathcal{O}(1)$ for the deflagration scenarios. The two peaks are located near $R^*$ and $\Delta R^*$, which support the findings of Refs.~\cite{Hindmarsh:2018ssm,Hindmarsh:2019ssm}. For deflagration scenarios, the GW spectra fall off more steeply as $k^{-8}$ in the ultraviolet. The peak appears near $R^*$ because $\Delta R^*$ is close to $R^*$ at this time, resulting in a single-peak feature. After dividing out the scaling factor $3(1+\omega)^2\overline{U}_f^4\Upsilon$ in Eq.~(\ref{eq:Omega_parameter}), we present the corresponding reduced spectra $\tilde{\Omega}_{gw}^{f}$ in {\it Supplemental Material}. Their peak amplitudes satisfy $\tilde{\Omega}_{gw}^{f}\sim \mathcal{O}(10^{-2})$ at the end of the PT ($\omega \simeq 1/3$), and are of the same order of magnitude as those observed in the flat-space simulations~\cite{Correia:2025qif,Hindmarsh:2017gnf,Hindmarsh:2015qta,Hindmarsh:2020hop}.

These slopes are instructive to compare with previous studies. The detonation ultraviolet tail $k^{-4}$ is close to the sound-shell-model prediction $k^{-3}$~\cite{Hindmarsh:2018ssm,Hindmarsh:2019ssm}, and matches the $k^{-4}$ high-frequency asymptote of the broken-power-law template $C(s)=s^{3}/(\tfrac{7}{4}+3s^{2})^{7/2}$ adopted for LISA forecasts~\cite{Caprini:2019egz,Hindmarsh:2019ssm}. The steeper $k^{-8}$ deflagration tail is qualitatively consistent with the steeper-than-$k^{-3}$ decay reported in~\cite{Hindmarsh:2017gnf}, though quantitatively steeper in our case, which likely arises from the stronger transition strength and the expanding background in our simulations. The steep $k^{7}$ segment between the two peaks is steeper than the $k^{1}$ plateau that the sound shell model predicts for very thin sound shells~\cite{Hindmarsh:2019ssm}.

\noindent{\it \bfseries Conclusion and discussion.}
We numerically study the first-order PT in an expanding Universe with the background scalar coupled to fluid framework. In this setup, the friction term is not a fixed constant but evolves dynamically during the PT. We observe that, in all simulations with PT strengths of $\alpha=0.15,\,0.38,\,0.60$, the sound wave contributions to the GW spectra dominate over the bubble-collision ones, and a larger $\alpha$ yields a smaller sound speed. In addition, rotational modes are generated in the fluid during the collisions, and their magnitude grows with the PT strength.
 
We quantify the amplitude of GW spectra from the sound wave contributions. The normalized peak amplitude $(\mathcal{H}_n R^*)^{-1}\Upsilon^{-1}\Omega_{gw}^{f}$ varies from $\mathcal{O}(10^{-2})$ up to $\mathcal{O}(1)$ for the detonation runs and proportional to the PT strength $\alpha$ and the duration of the PT  $((\beta/H)^{-1})$. After accounting for the growth factor $\Upsilon$, the amplitude $\tilde{\Omega}_{gw}^{f}$ reduces to $\mathcal{O}(10^{-2})$ for both scenarios. 
 This aligns with flat-spacetime lattice simulations, where $ \tilde{\Omega}_{gw} $ reaches $ \mathcal{O}(10^{-2}) $ for weak transitions ( $ \alpha \sim 10^{-2} $ )~\cite{Hindmarsh:2015qta,Caprini:2019egz} and strong transitions ( $ \alpha \approx 0.5 $ )~\cite{Correia:2025qif}. For detonation scenarios, the sound wave spectrum shows a double-peak structure, and only a single peak remains for deflagrations. The bubble-collision spectra, shown in the {\it Supplemental Material}, fall off as $k^{-1}$ in the ultraviolet for detonations and show a single peak with a $k^{-4}$ tail for deflagrations. These features provide useful diagnostics for the GW signals to be searched for by LISA, Taiji and TianQin~\cite{LISA:2017pwj,Caprini:2019egz,TianQin:2015yph}.

The technique developed here can be generalized to precisely investigate the dark matter particle production during the bubble collisions and the ossicillation stage~\cite{Azatov:2021ifm,Ai:2024ikj,Shakya:2023kjf,Mansour:2023fwj,Falkowski:2012fb,Baldes:2023cih} and the assiciated GWs production~\cite{Inomata:2024rkt}, and also the density perturbation and primordial black hole formation possibility in the presence of thermal plasma~\cite{Zou:2025sow}. 

\noindent{\it \bfseries Acknowledgments.}
We are grateful to Daniel Cutting, Huai-Ke Guo, Shao-Jiang Wang, David J. Weir, Chi Tian, and Zhi-Qing Zhu for helpful discussions.
This work is supported by the National Natural Science Foundation of China (NSFC) under Grants Nos. 12675130 and 12547101.
L.B. also acknowledges Chongqing Natural Science Foundation under Grant
No. CSTB2024NSCQ-JQX0022.

\bibliography{references}

\clearpage

\onecolumngrid
\begin{center}
  \textbf{\large Supplemental Material}\\[.2cm]
\end{center}

In this Supplemental Material, we present (i) the evolution of the phase transition parameters and slices of the field, temperature, fluid velocity, and fluid vorticity during the phase transition, and (ii) the time evolution of the sound wave spectra, the end-state spectra of both the sound wave and the bubble-collision contributions, and a summary table of the parameter values and the resulting GW amplitudes.

\section{The evolution of the phase transition parameters and the slices during the PT}
\begin{figure*}[!htbp]
    \centering
    \includegraphics[width=0.32\textwidth,clip]{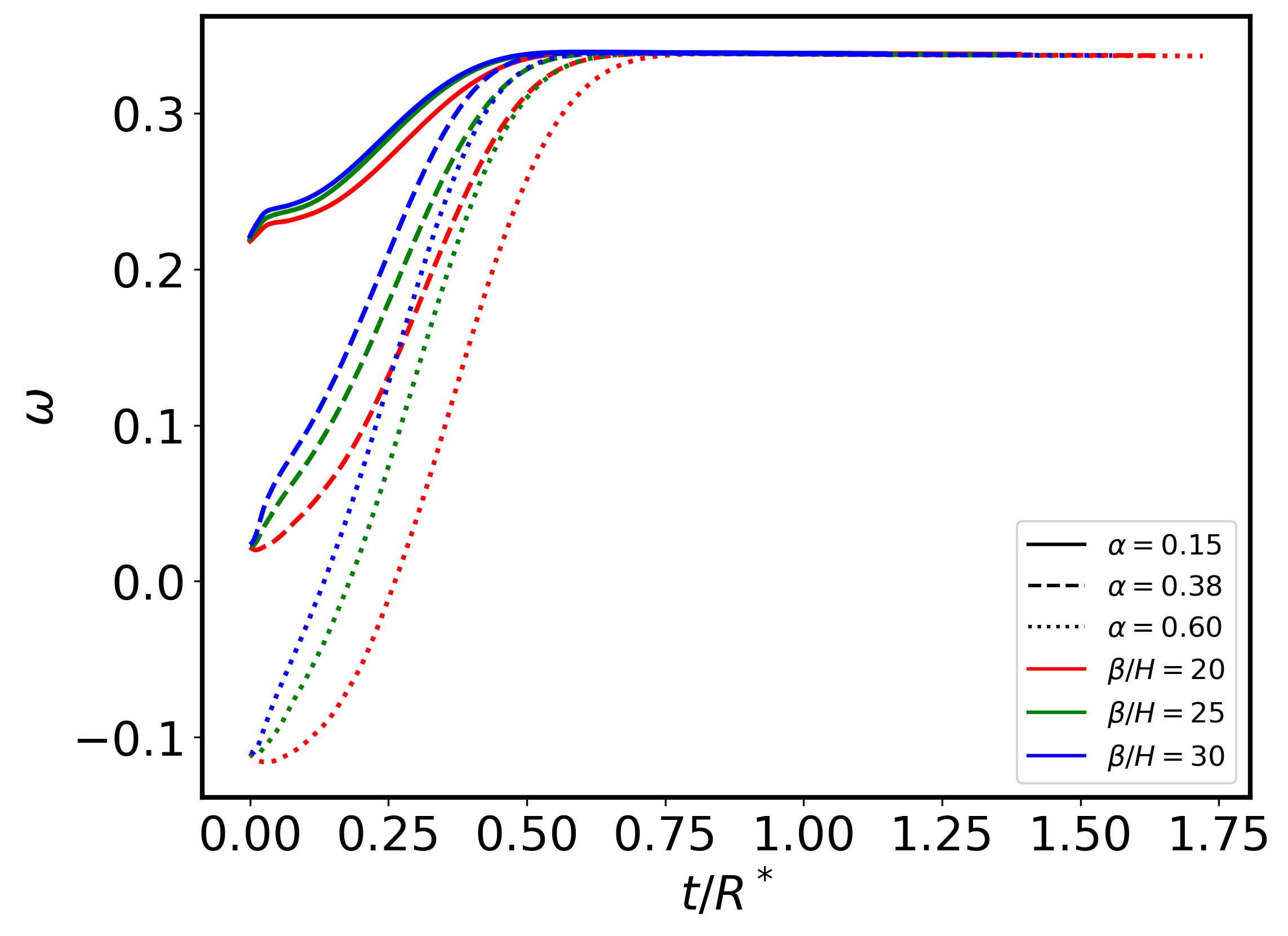}
    \includegraphics[width=0.32\textwidth,clip]{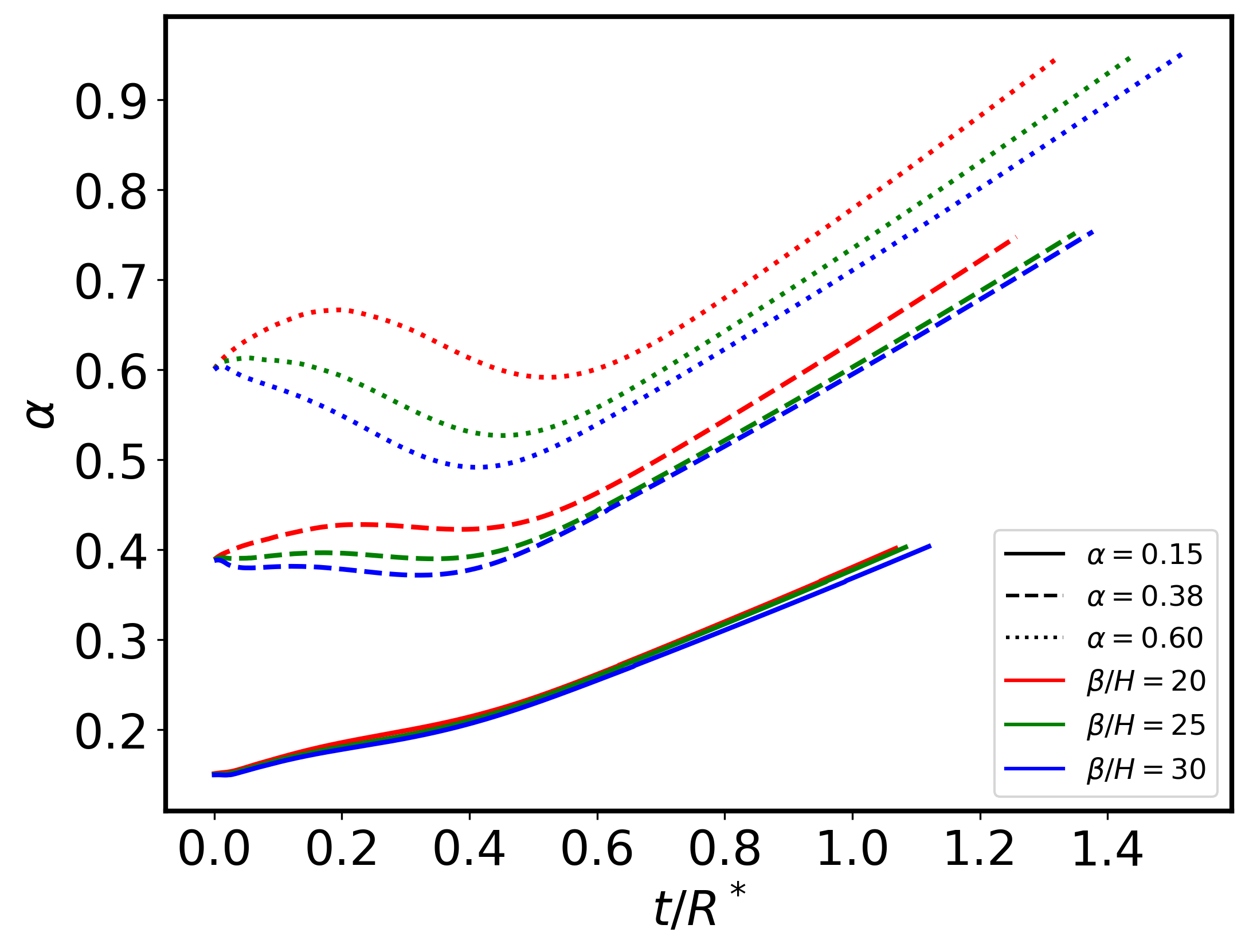}
    \includegraphics[width=0.32\textwidth,clip]{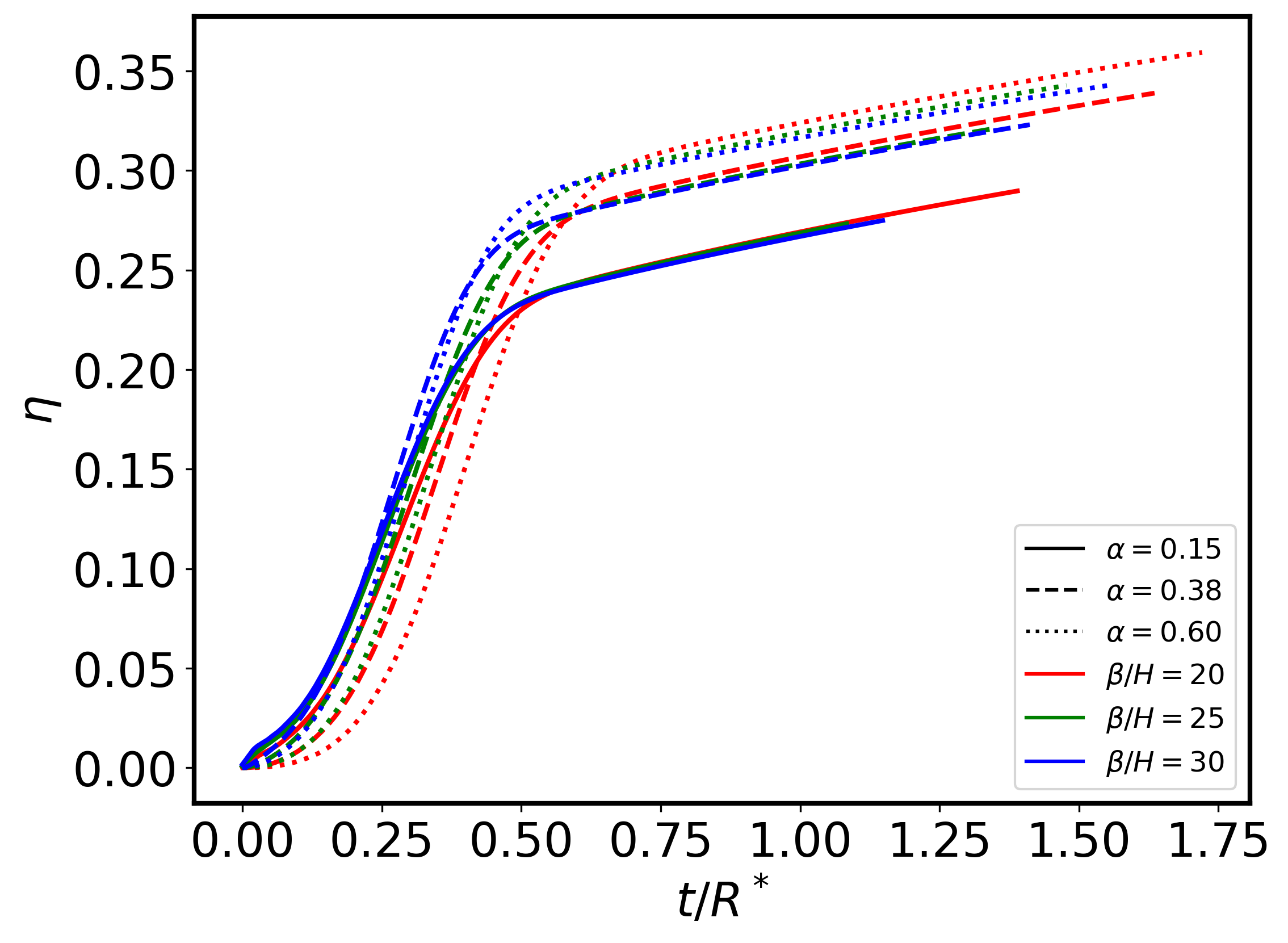}\\
    \includegraphics[width=0.32\textwidth,clip]{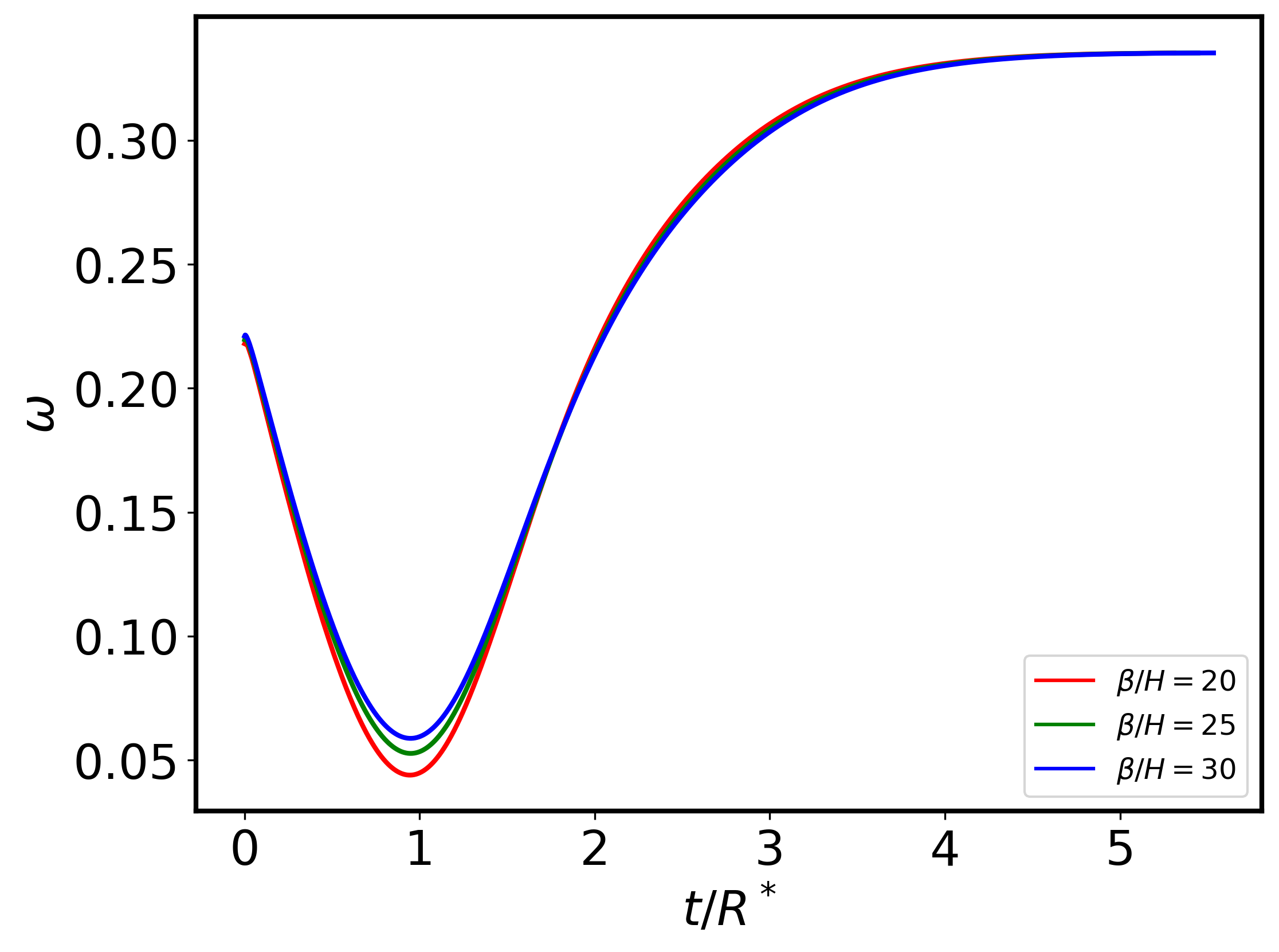}
    \includegraphics[width=0.32\textwidth,clip]{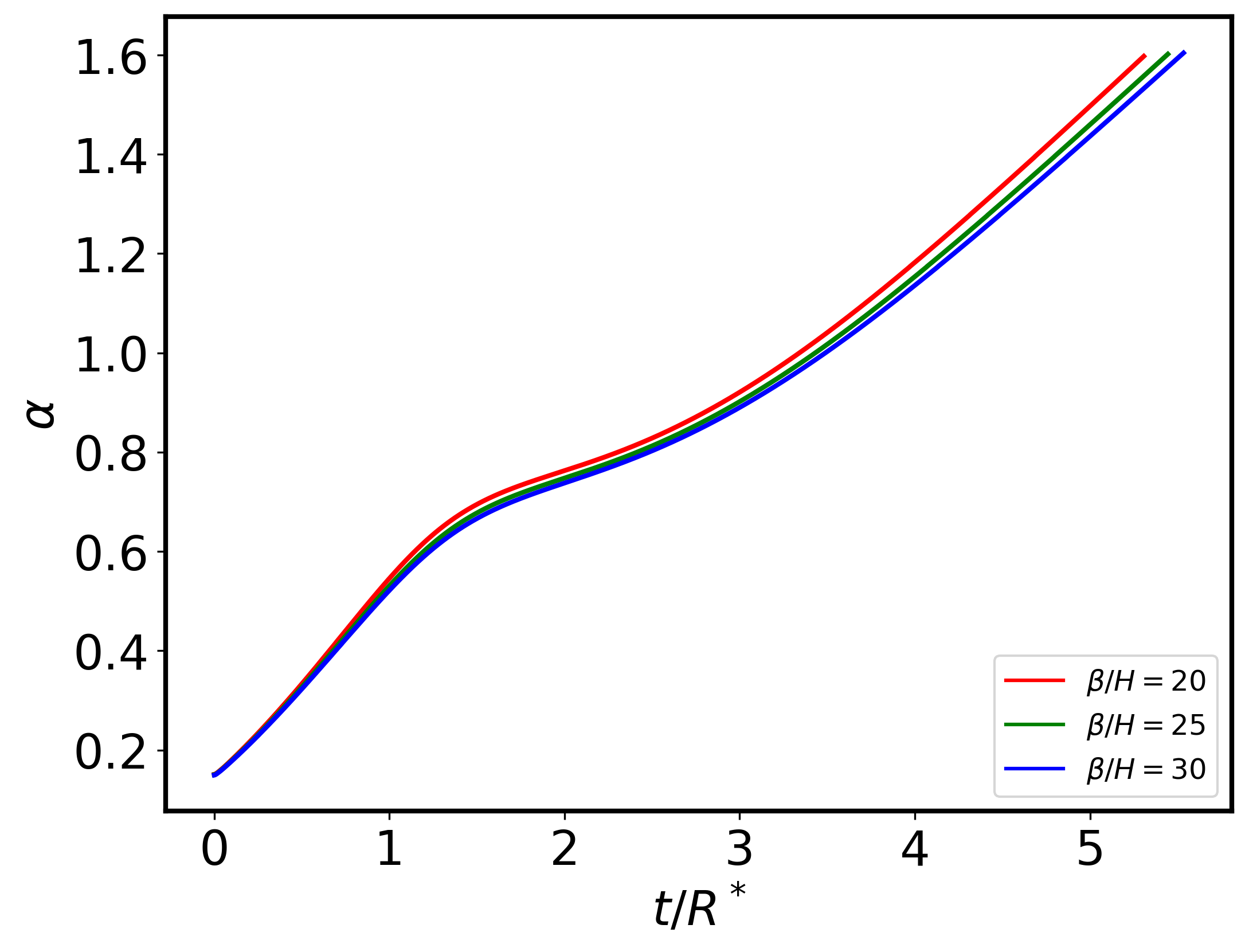}
    \includegraphics[width=0.32\textwidth,clip]{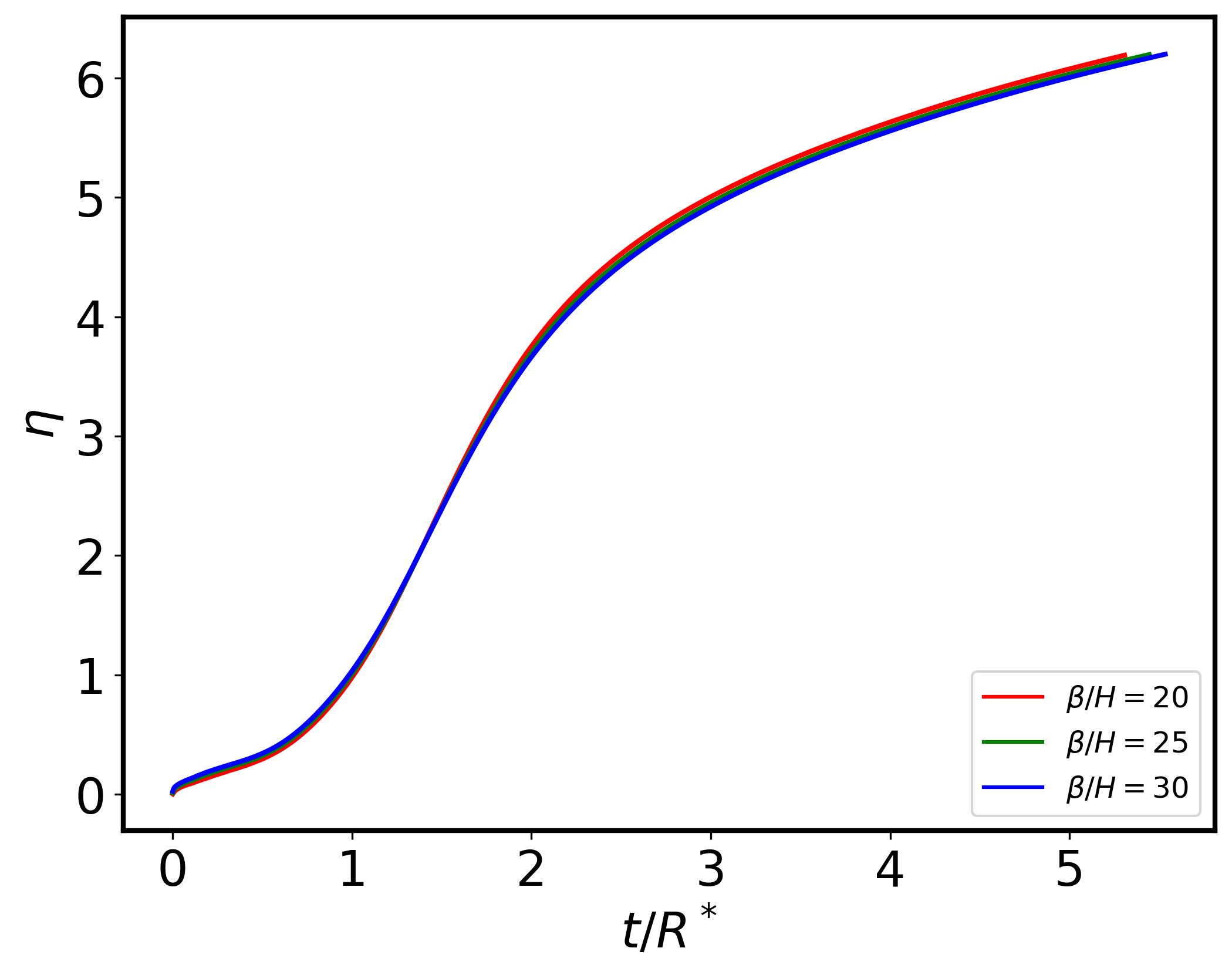}
    \caption{Evolution of the equation-of-state parameter $\omega $, the phase transition strength $\alpha$ and the friction parameter  $\eta=\tilde{\eta}\frac{\phi^2}{T}$ in the expanding Universe, for detonations (top row) and deflagrations (bottom row).}
\label{fig:EOS_eta}
\end{figure*}

Fig.~\ref{fig:EOS_eta} shows the evolution of the equation-of-state parameter $\omega$ (left), the phase transition strength $\alpha$ (middle), and the friction coupling $\eta$ (right), for detonations (top row) and deflagrations (bottom row). The initial value of $\omega$ is smaller for a larger $\alpha_0$: a stronger transition stores a larger fraction of the total energy in the form of vacuum energy, so the equation of state deviates further from that of pure radiation. As time evolves, simulations with larger $\beta/H$ evolve to $\omega = 1/3$ more rapidly. The middle panel shows that cosmic expansion lowers the temperature, and the resulting change in the entropy density difference between the two phases causes $\alpha$ to grow during the evolution. As shown in the right panel, the gradual increase of the field value $\phi$ and the concurrent drop in temperature during the evolution drive the gradual rise of $\eta$, which is larger for a larger $\alpha$ and depends only weakly on $\beta/H$.

\begin{figure*}[!htbp]
    \centering
    \includegraphics[width=0.32\textwidth,clip]{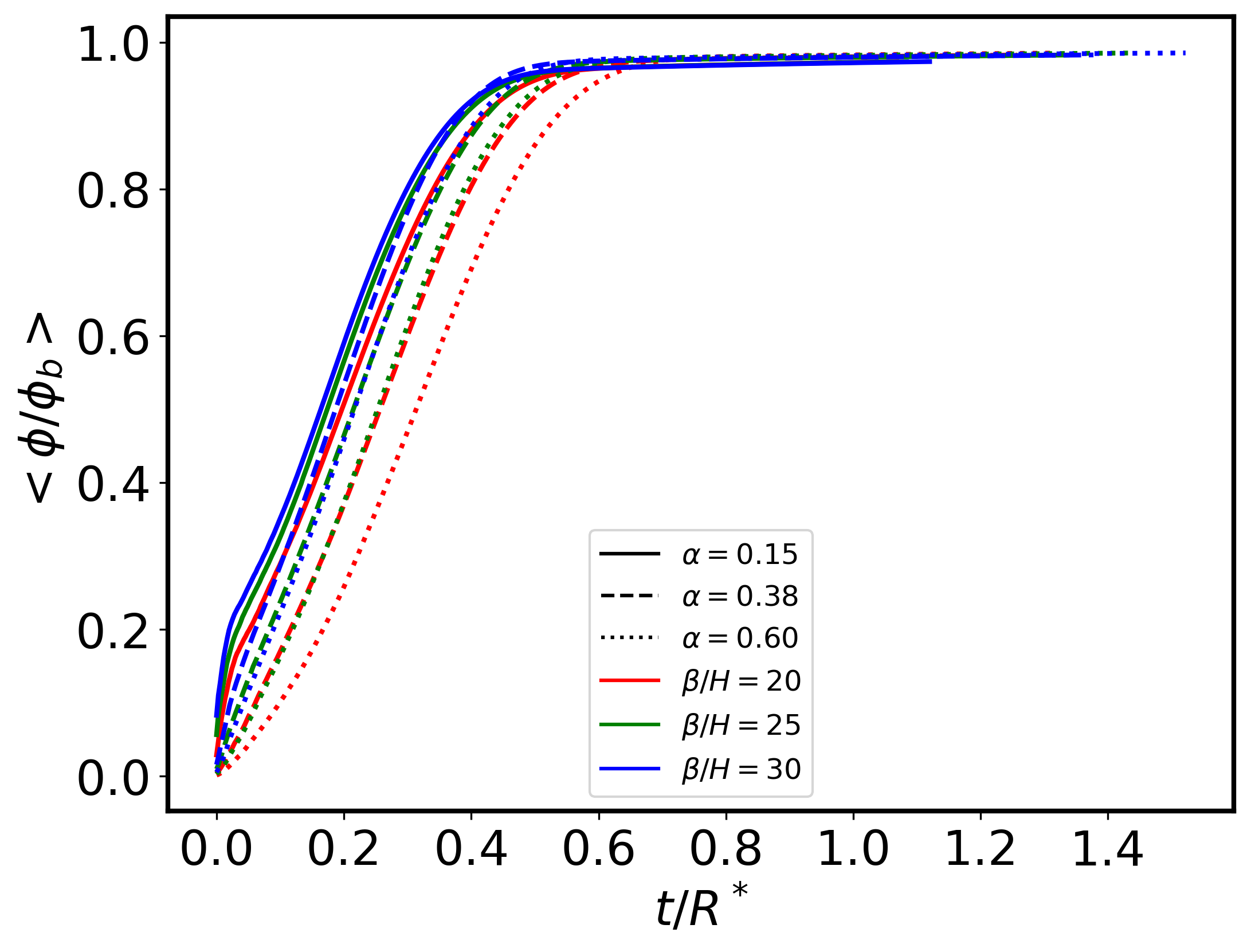}
    \includegraphics[width=0.32\textwidth,clip]{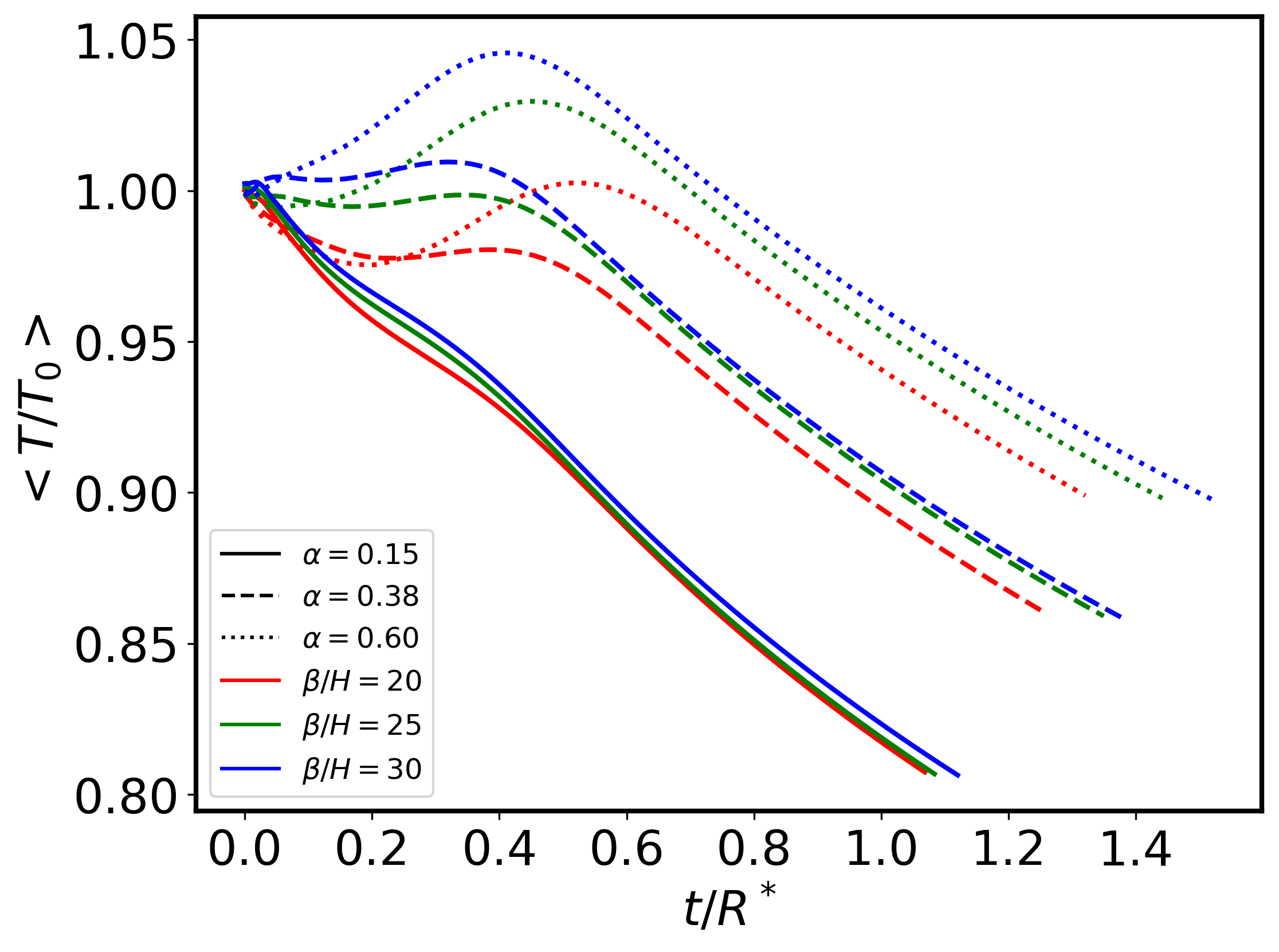}
    \includegraphics[width=0.32\textwidth,clip]{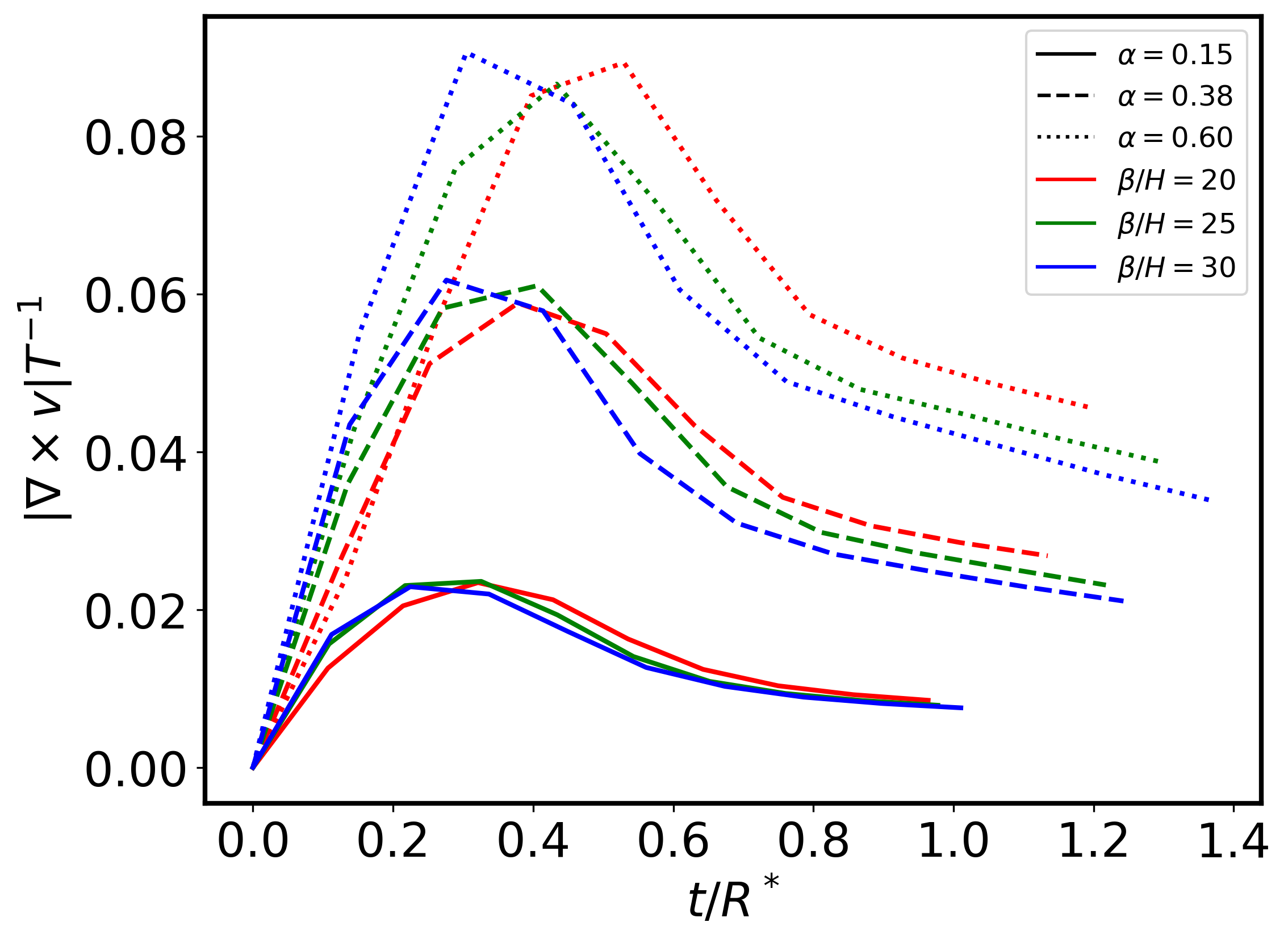}\\
    \includegraphics[width=0.32\textwidth,clip]{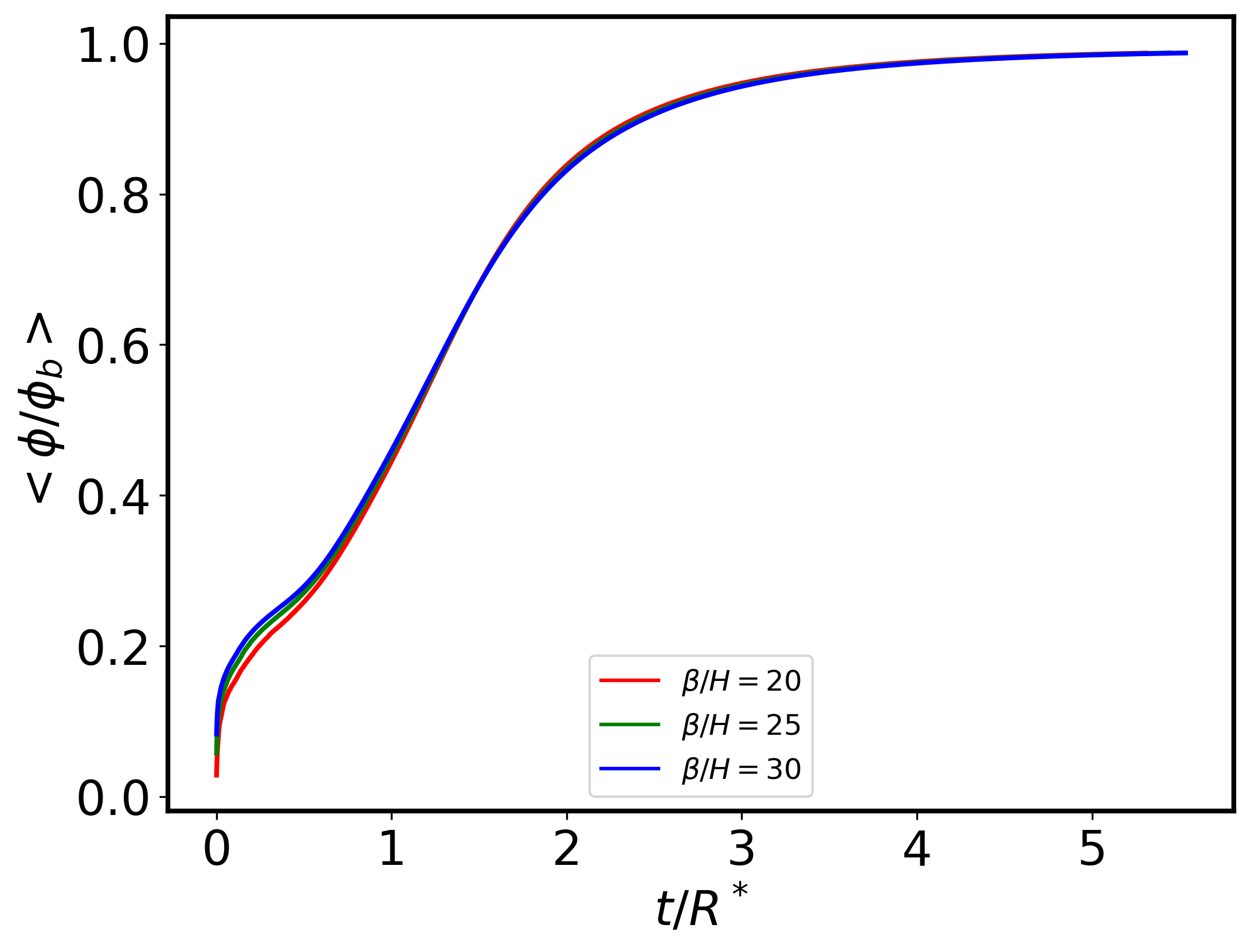}
    \includegraphics[width=0.32\textwidth,clip]{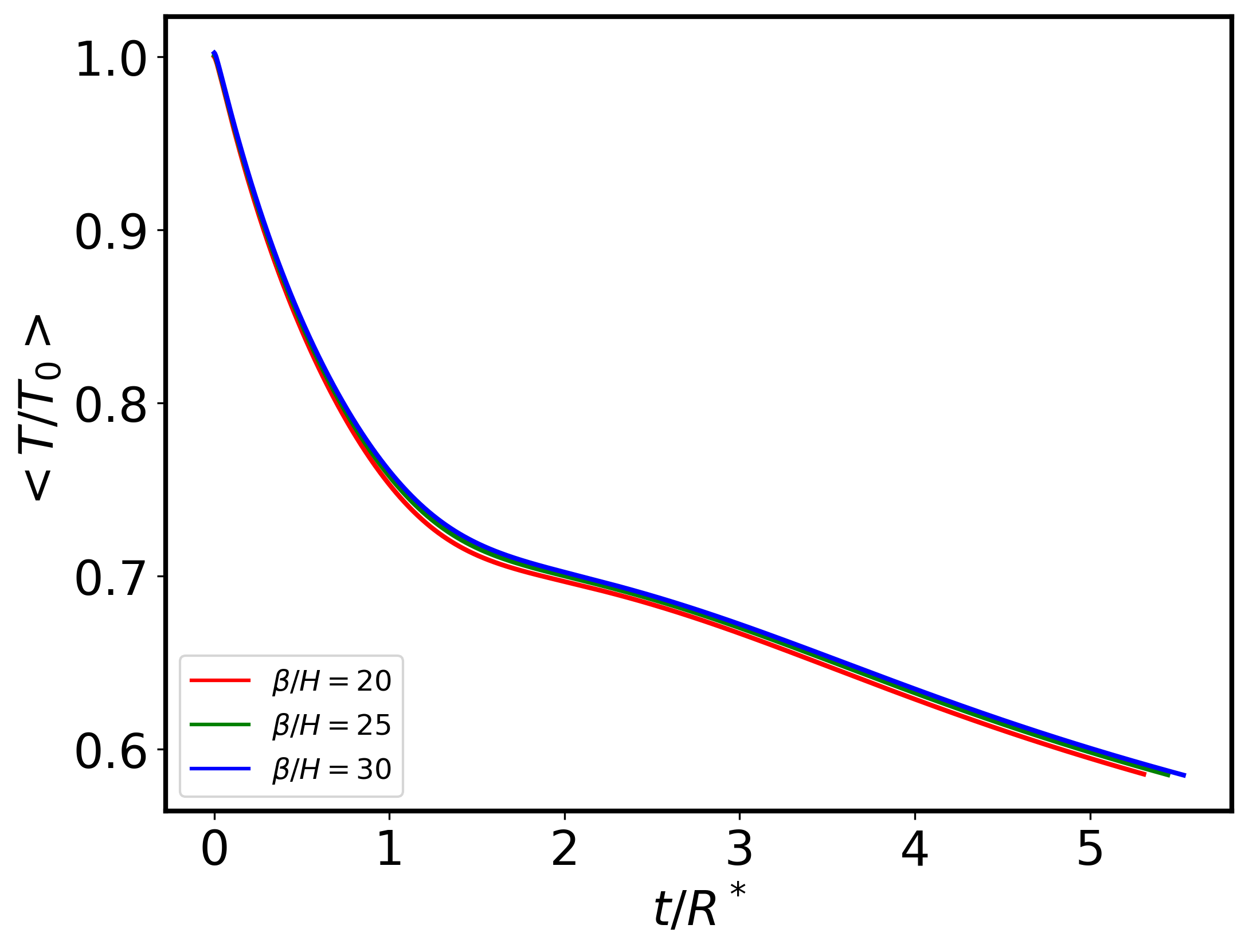}
    \includegraphics[width=0.32\textwidth,clip]{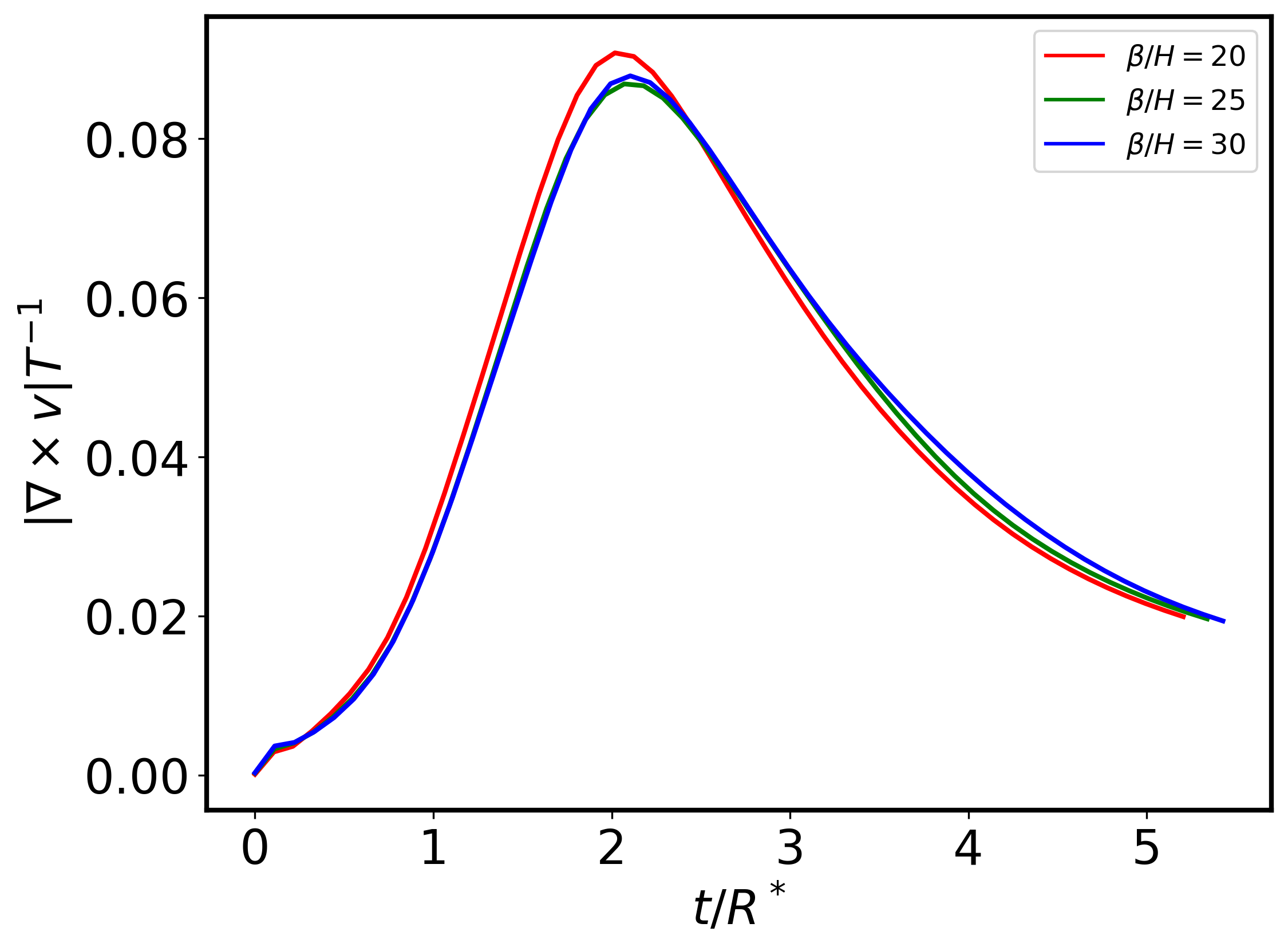}
    \caption{The mean field evolution $\phi/\phi_b$, the mean temperature evolution $T/T_0$ in the expanding Universe, and the average magnitude of the velocity curl for detonations (top row) and deflagrations (bottom row).}
\label{fig:field_curl}
\end{figure*}

Fig.~\ref{fig:field_curl} shows the evolution of the mean field $\phi/\phi_b$ (left), the mean temperature $T/T_0$ (middle), and the average magnitude of the velocity curl (right). For the mean field, a larger $\alpha$ leads to a slower field evolution, while a larger $\beta/H$ leads to a faster one. For detonations, the temperature is mainly concentrated on the bubble-collision surfaces. When $\alpha$ is large (0.38, 0.60), the mean temperature rises around $t/R_* = 0.5$: this is the period of bubble collisions, and for a strong transition such as $\alpha=0.60$ a large amount of vacuum energy is released into the plasma, leading to substantial reheating of the Universe at a rate faster than the universe expansion dilutes it. For $\alpha = 0.15$, the conversion is slower than the dilution, so this rise does not appear. Afterwards, in all cases the temperature decreases with the universe expansion. For deflagrations, during the nucleation stage the temperature is mainly concentrated at the bubble centers; since the wall speed is low, the energy gradually flows into the radiation energy of the fluid, raising the fluid temperature, and the temperature drops more rapidly before the collision and more slowly after it. We note that the temperature evolution is complementary to the evolution of the equation of state $\omega$ shown in Fig.~\ref{fig:EOS_eta}. In addition, we examine the curl of the velocity field. Its magnitude grows with the PT strength $\alpha$, a stronger transition transfers more energy into the fluid, so the fluid moves with a larger velocity.
\begin{figure*}[!htp]
    \centering
	\includegraphics[width=0.24\textwidth,clip]{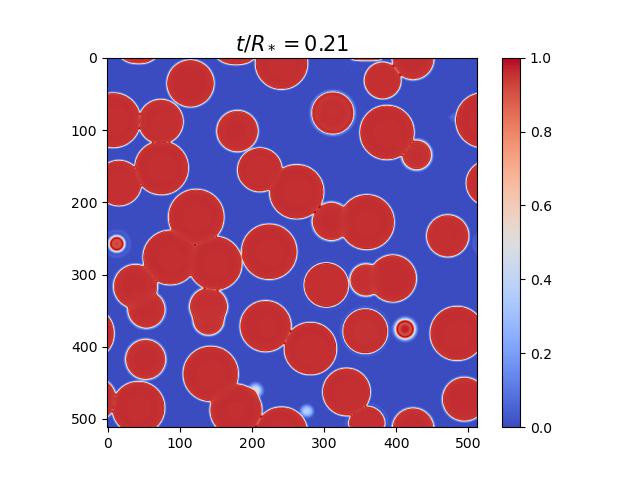}
	\includegraphics[width=0.24\textwidth,clip]{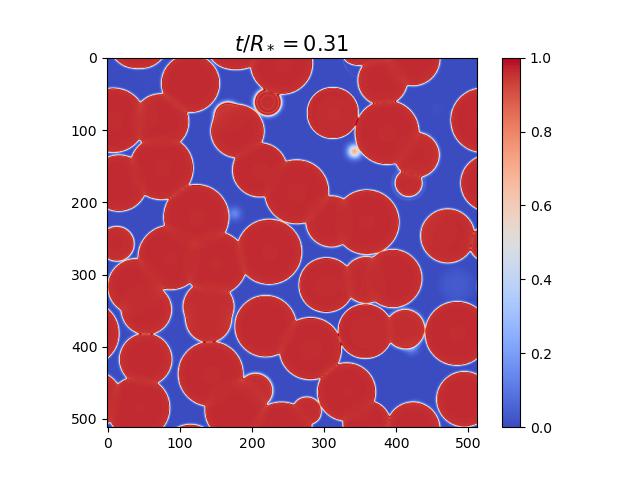}
	\includegraphics[width=0.24\textwidth,clip]{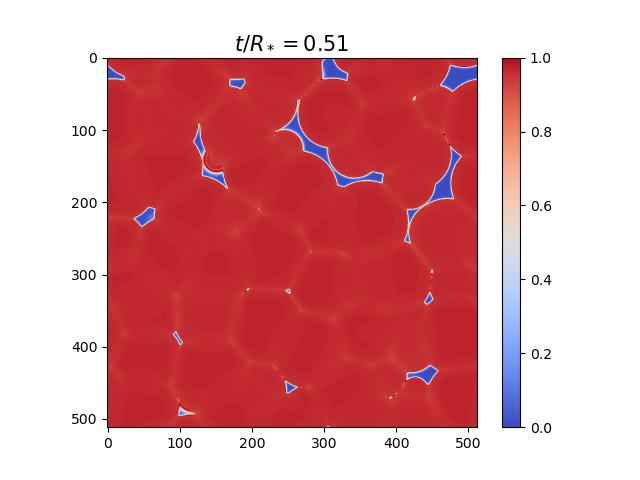}
	\includegraphics[width=0.24\textwidth,clip]{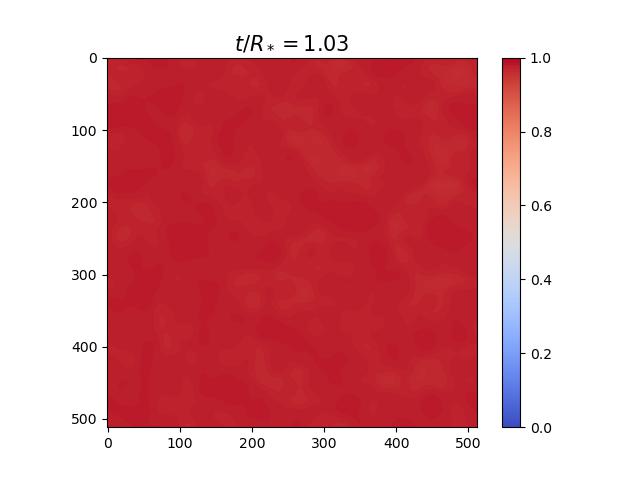}\\
	\includegraphics[width=0.24\textwidth,clip]{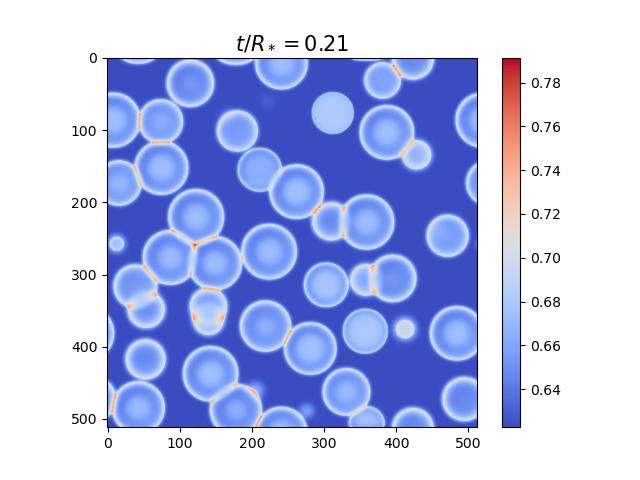}
	\includegraphics[width=0.24\textwidth,clip]{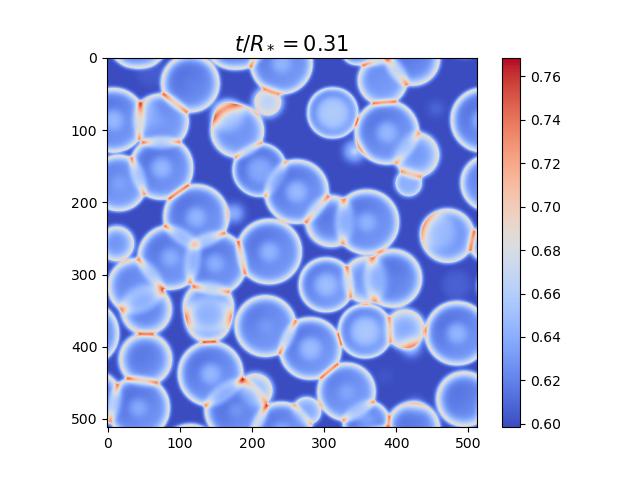}
	\includegraphics[width=0.24\textwidth,clip]{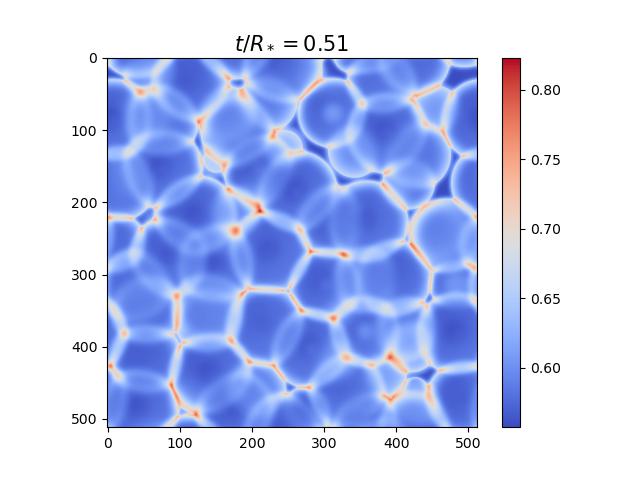}
	\includegraphics[width=0.24\textwidth,clip]{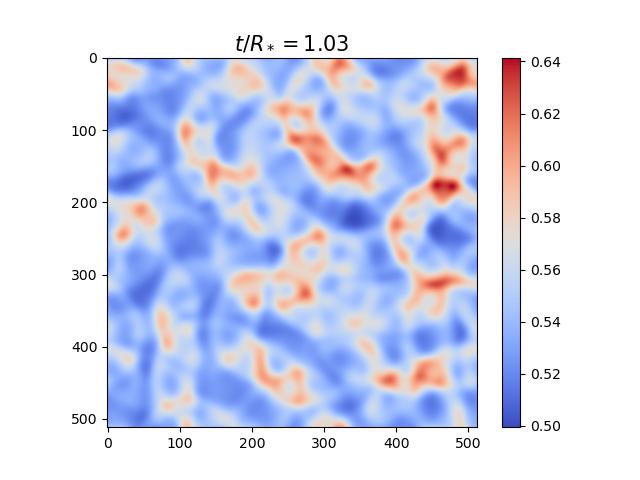}\\
	\includegraphics[width=0.24\textwidth,clip]{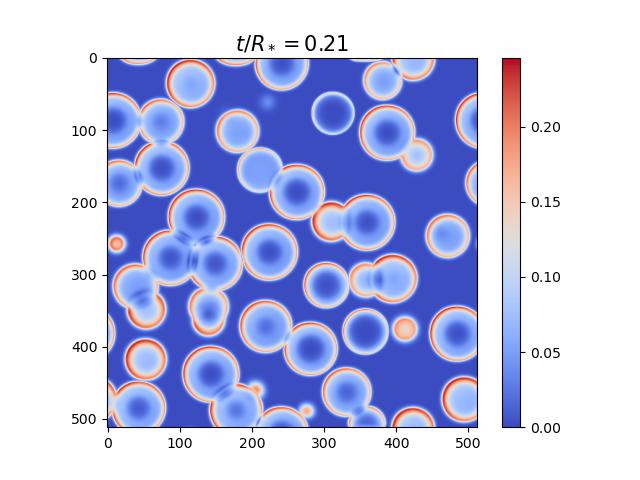}
	\includegraphics[width=0.24\textwidth,clip]{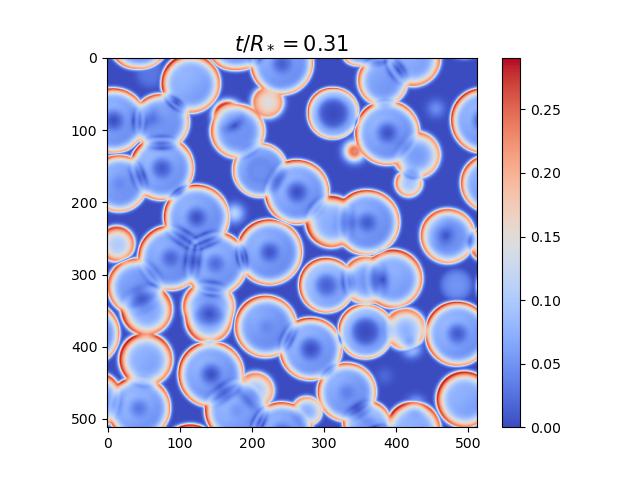}
	\includegraphics[width=0.24\textwidth,clip]{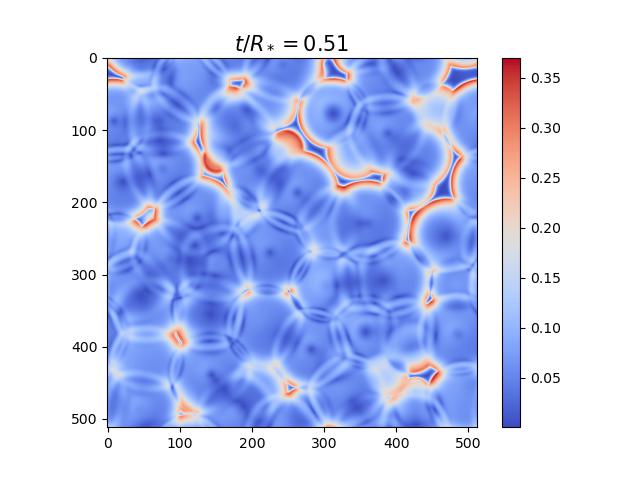}
	\includegraphics[width=0.24\textwidth,clip]{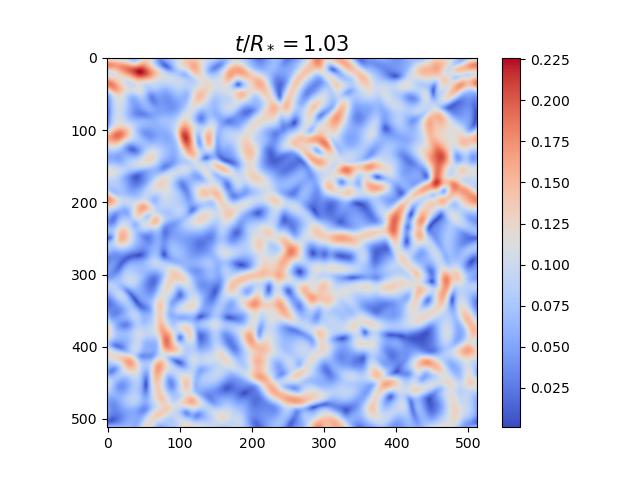}\\
	\includegraphics[width=0.24\textwidth,clip]{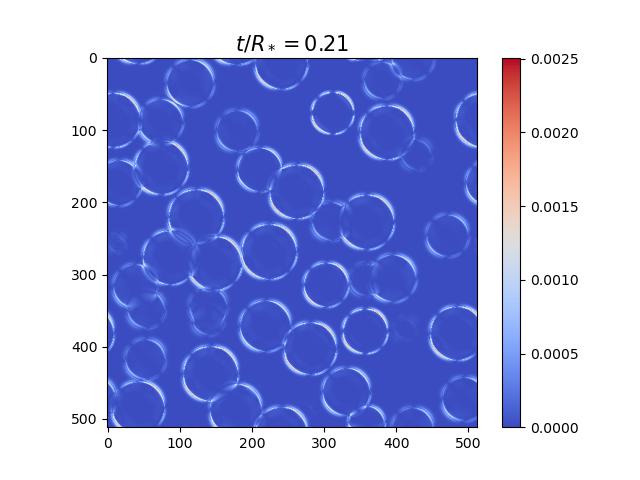}
	\includegraphics[width=0.24\textwidth,clip]{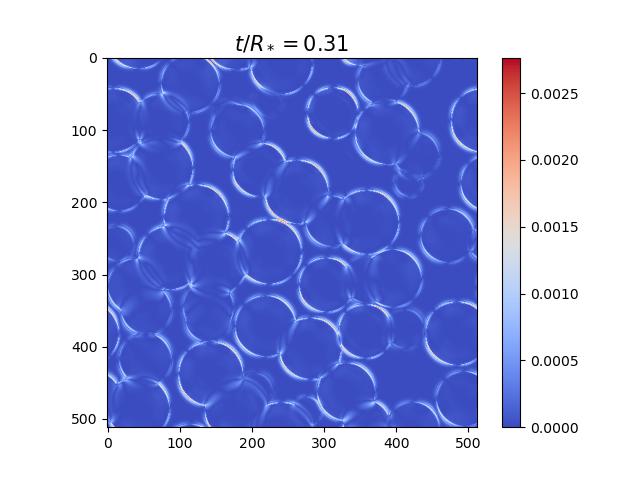}
	\includegraphics[width=0.24\textwidth,clip]{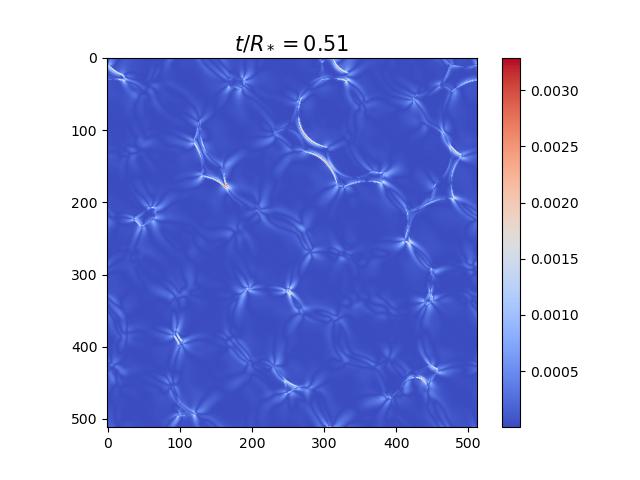}
	\includegraphics[width=0.24\textwidth,clip]{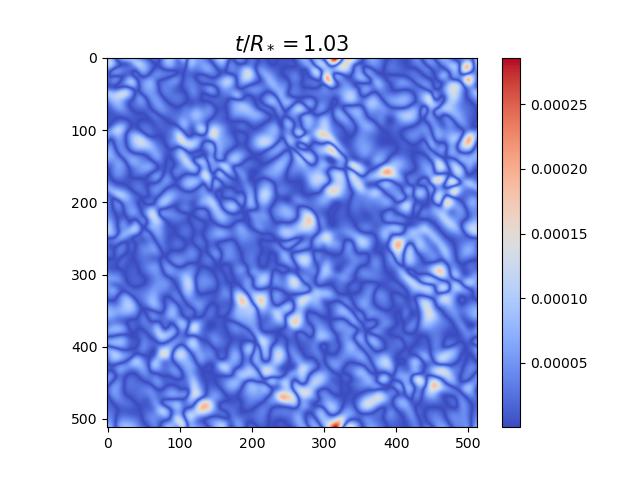}
	\caption{From top to bottom, the slices are the distribution of field $\phi/\phi_b$, temperature $T/T_0$, magnitude of fluid 3-velocity and magnitude of the curl of fluid 3-velocity. They are the results for the detonation branch with $\alpha_0 = 0.15$, $\beta/H=20$ and all locate on the $(100dx,y,z)$ plane.}
\label{fig:slice_det} 
\end{figure*}

\begin{figure*}[!htp]
    \centering
	\includegraphics[width=0.24\textwidth,clip]{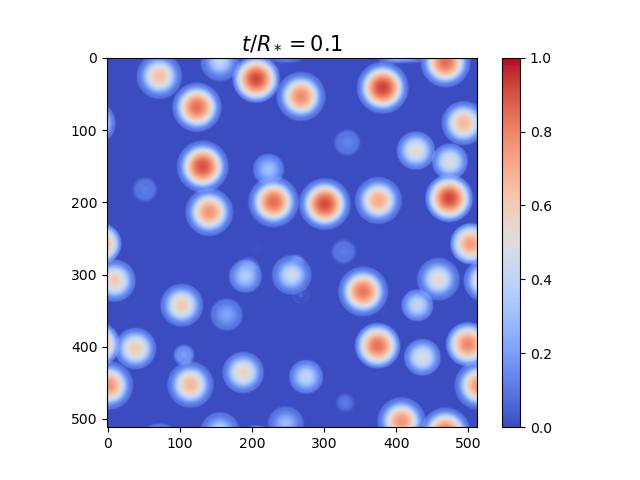}
	\includegraphics[width=0.24\textwidth,clip]{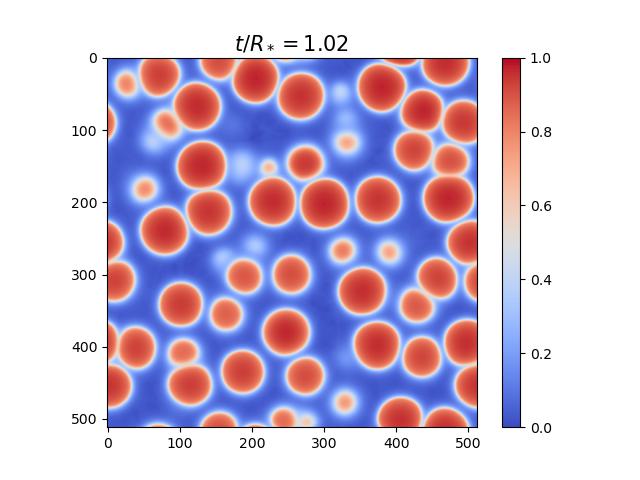}
	\includegraphics[width=0.24\textwidth,clip]{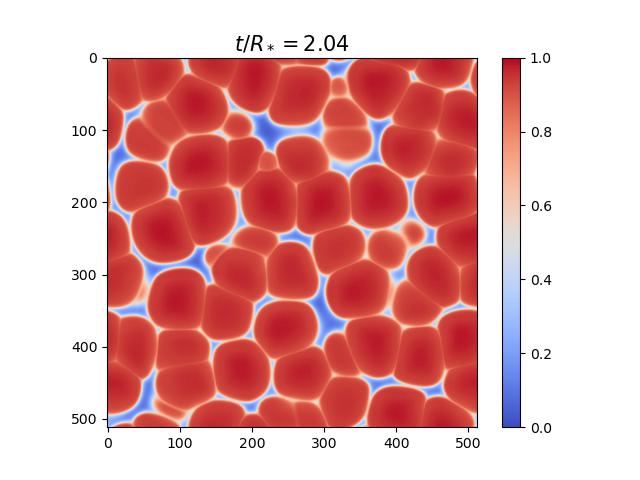}
	\includegraphics[width=0.24\textwidth,clip]{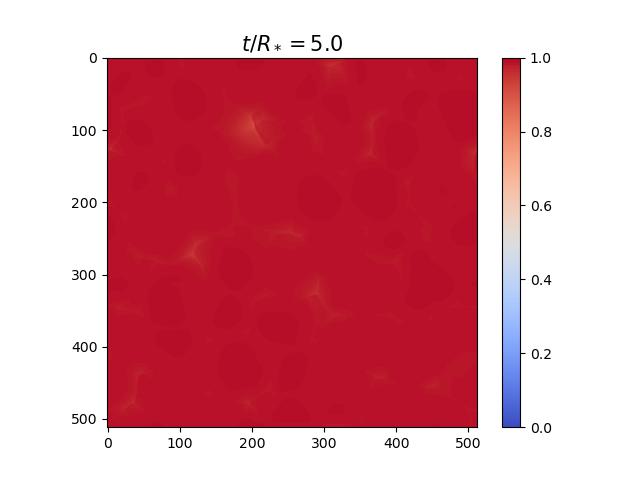}\\
	\includegraphics[width=0.24\textwidth,clip]{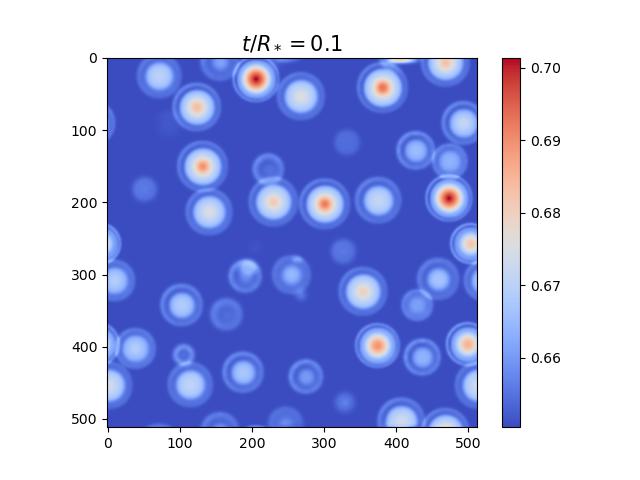}
	\includegraphics[width=0.24\textwidth,clip]{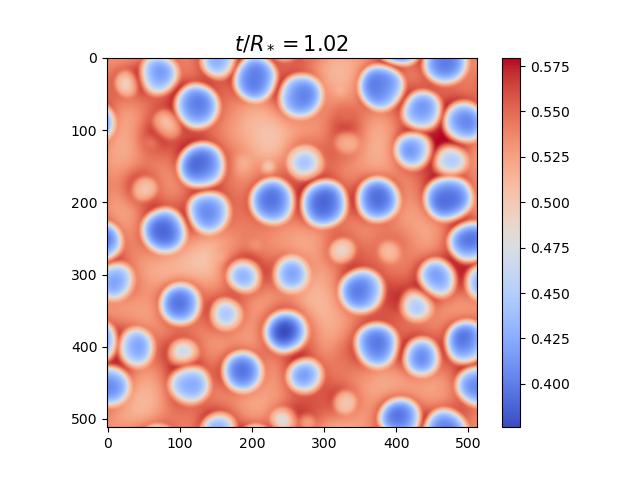}
	\includegraphics[width=0.24\textwidth,clip]{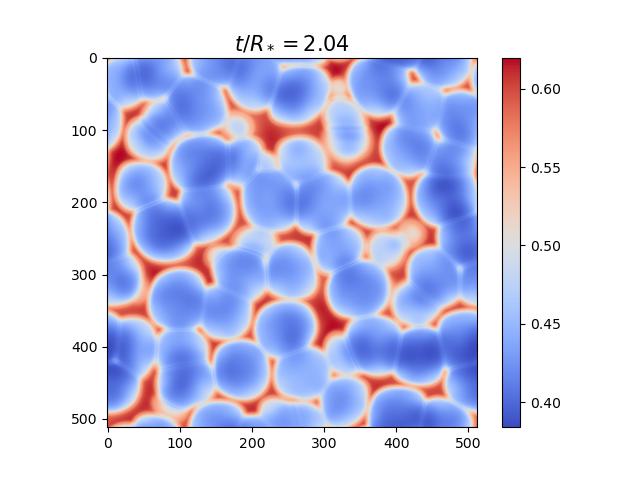}
	\includegraphics[width=0.24\textwidth,clip]{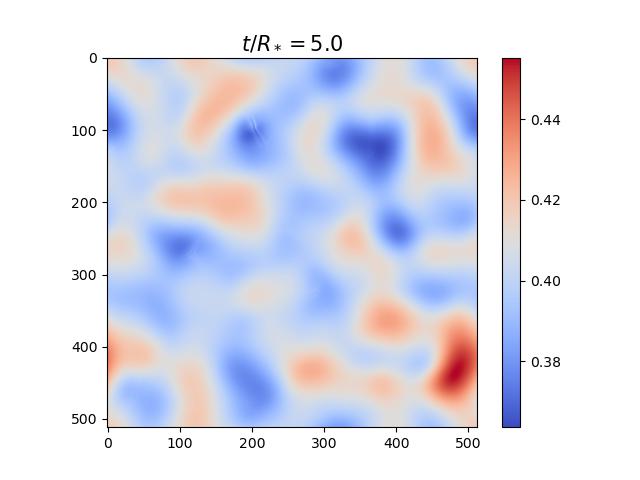}\\
	\includegraphics[width=0.24\textwidth,clip]{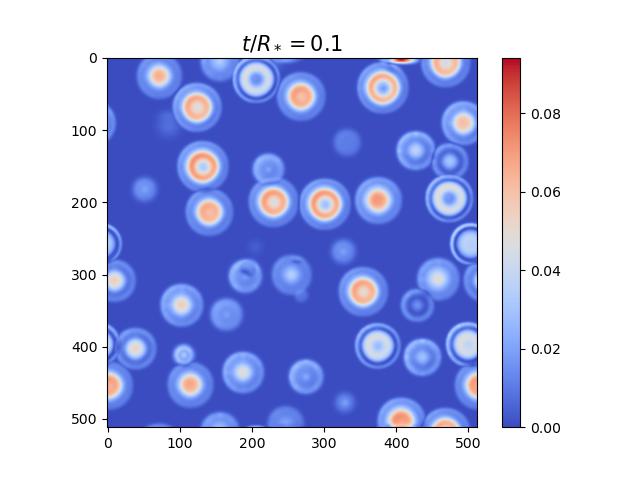}
	\includegraphics[width=0.24\textwidth,clip]{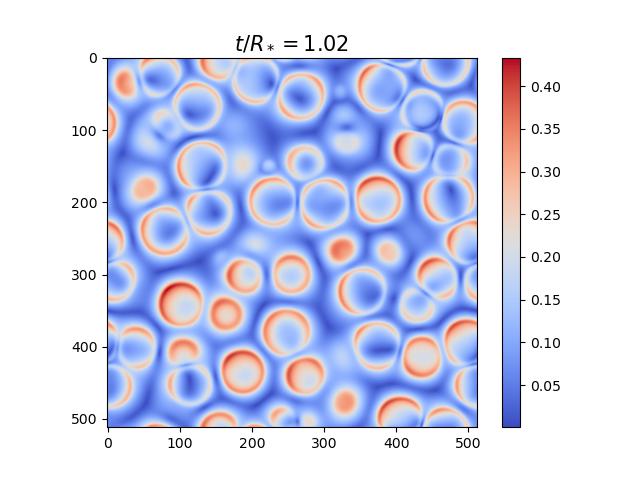}
	\includegraphics[width=0.24\textwidth,clip]{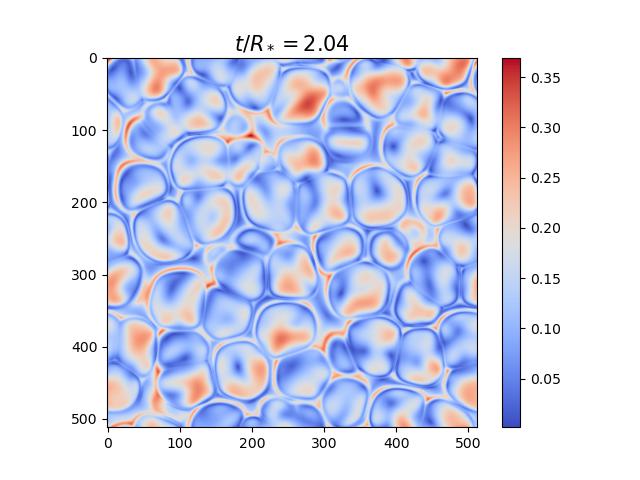}
	\includegraphics[width=0.24\textwidth,clip]{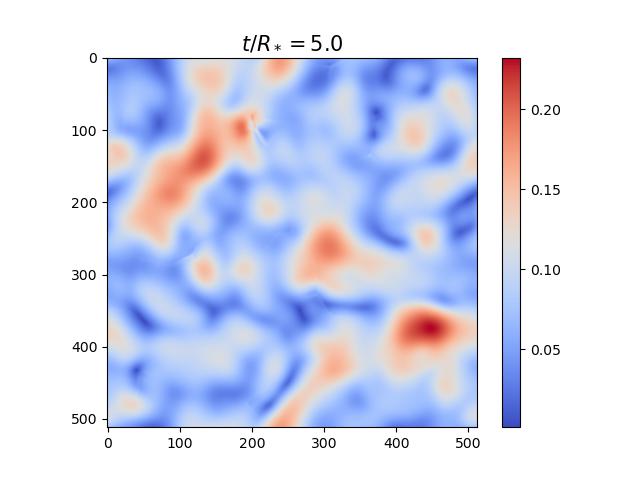}\\
	\includegraphics[width=0.24\textwidth,clip]{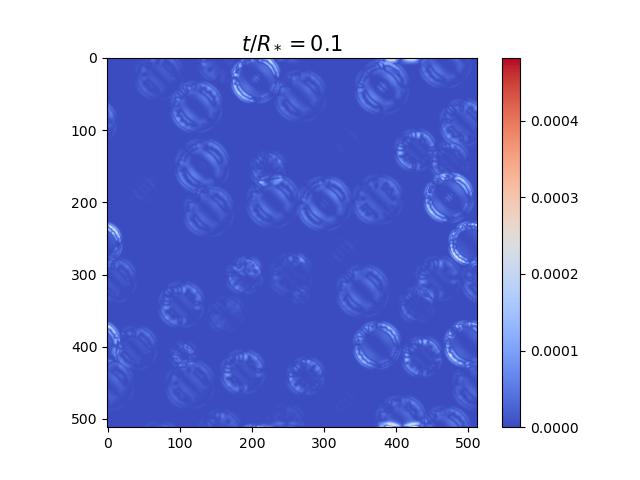}
	\includegraphics[width=0.24\textwidth,clip]{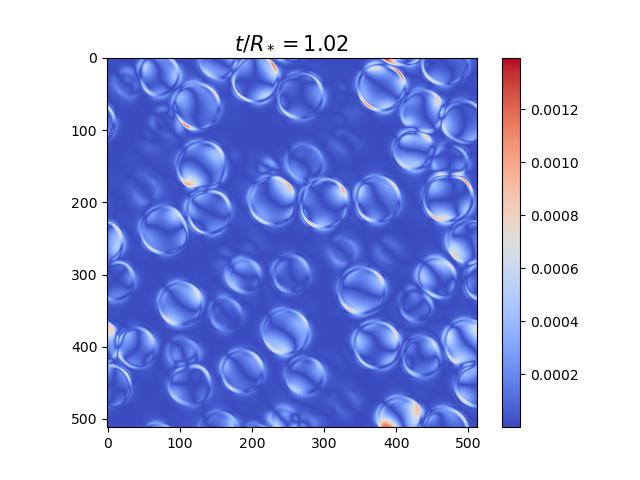}
	\includegraphics[width=0.24\textwidth,clip]{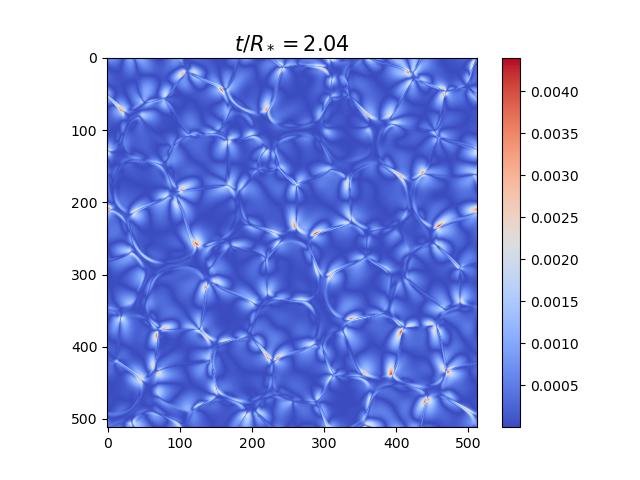}
	\includegraphics[width=0.24\textwidth,clip]{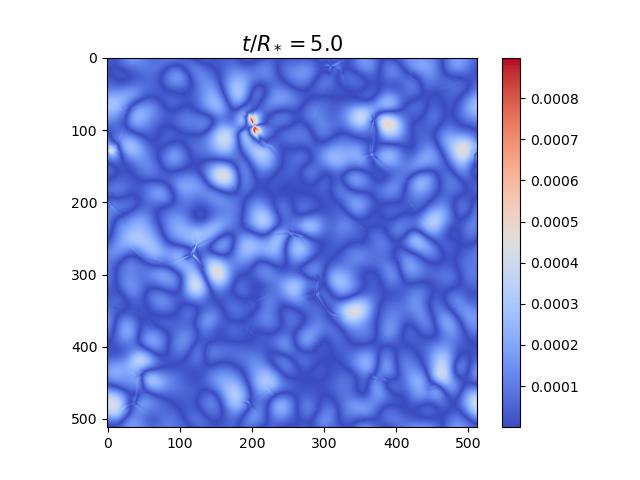}\\
	\caption{From top to bottom, the slices are the distribution of field $\phi/\phi_b$, temperature $T/T_0$, magnitude of fluid 3-velocity and magnitude of the curl of fluid 3-velocity. They are the results for the deflagration branch with $\alpha_0 = 0.15$, $\beta/H=20$, and all locate on the $(100dx,y,z)$ plane.}
\label{fig:slice_def} 
\end{figure*}

To visualize bubble expansion and the phase transition process, we present slices of the field, temperature, velocity magnitude, and velocity curl magnitude in the expanding Universe, for the detonation branch (Fig.~\ref{fig:slice_det}) and the deflagration branch (Fig.~\ref{fig:slice_def}), both with $\alpha_0 = 0.15$ and $\beta/H = 20$. In the detonation slices, the fluid remains nearly at rest outside the bubbles, with velocity concentrated primarily within the thin bubble walls. In the deflagration slices, by contrast, the fluid is set into broader motion, and the transition takes considerably longer to complete. At the end of the phase transition, bubbles collide and merge, and the deviation of bubbles from perfect spherical symmetry arises due to the influence of the fluid dynamics. This phenomenon is more pronounced in the deflagration regime.

\section{The GW spectra during and at the end of the phase transition}
\begin{figure*}[!htbp]
    \centering
   	\includegraphics[width=0.32\textwidth]{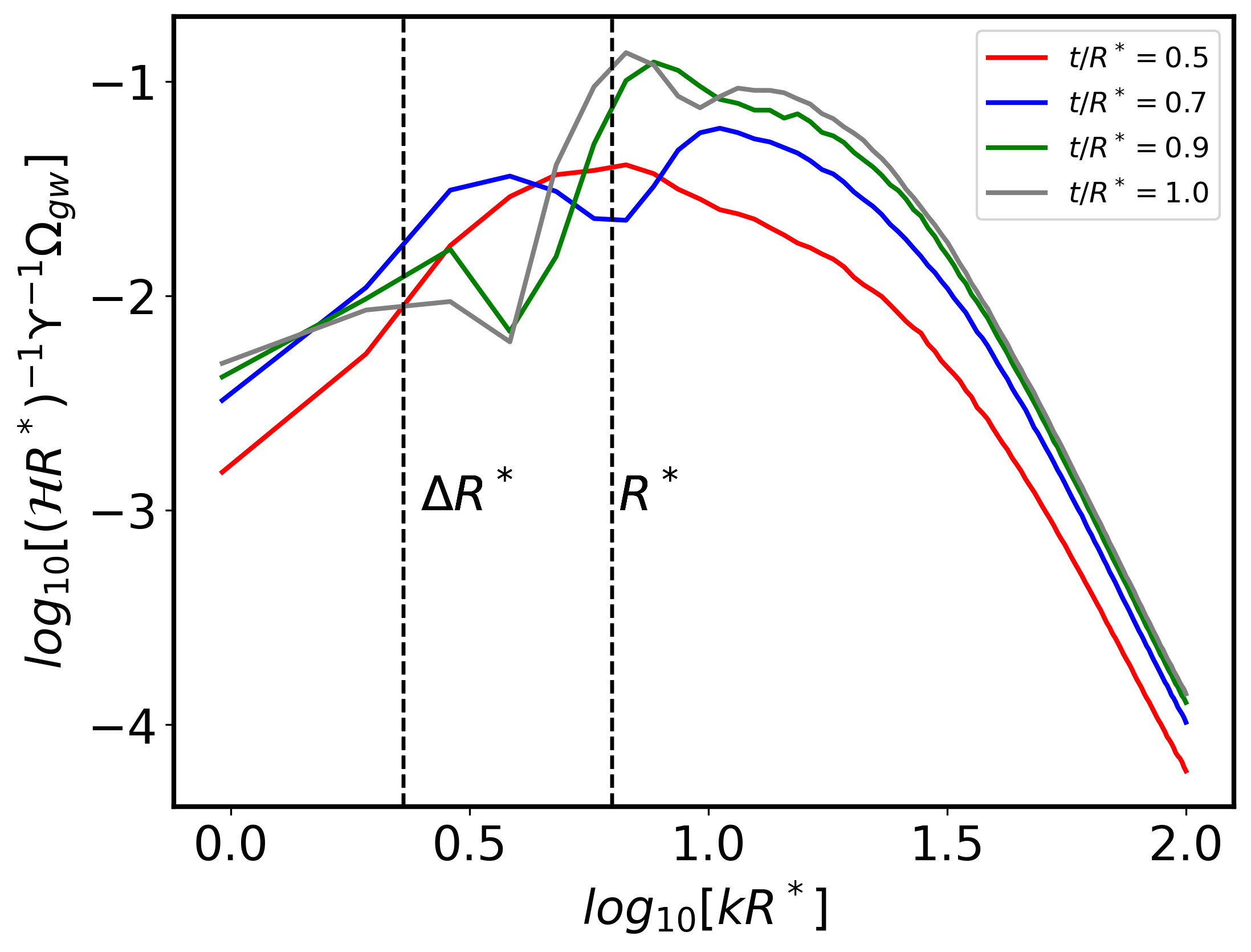}
   	\includegraphics[width=0.32\textwidth,clip]{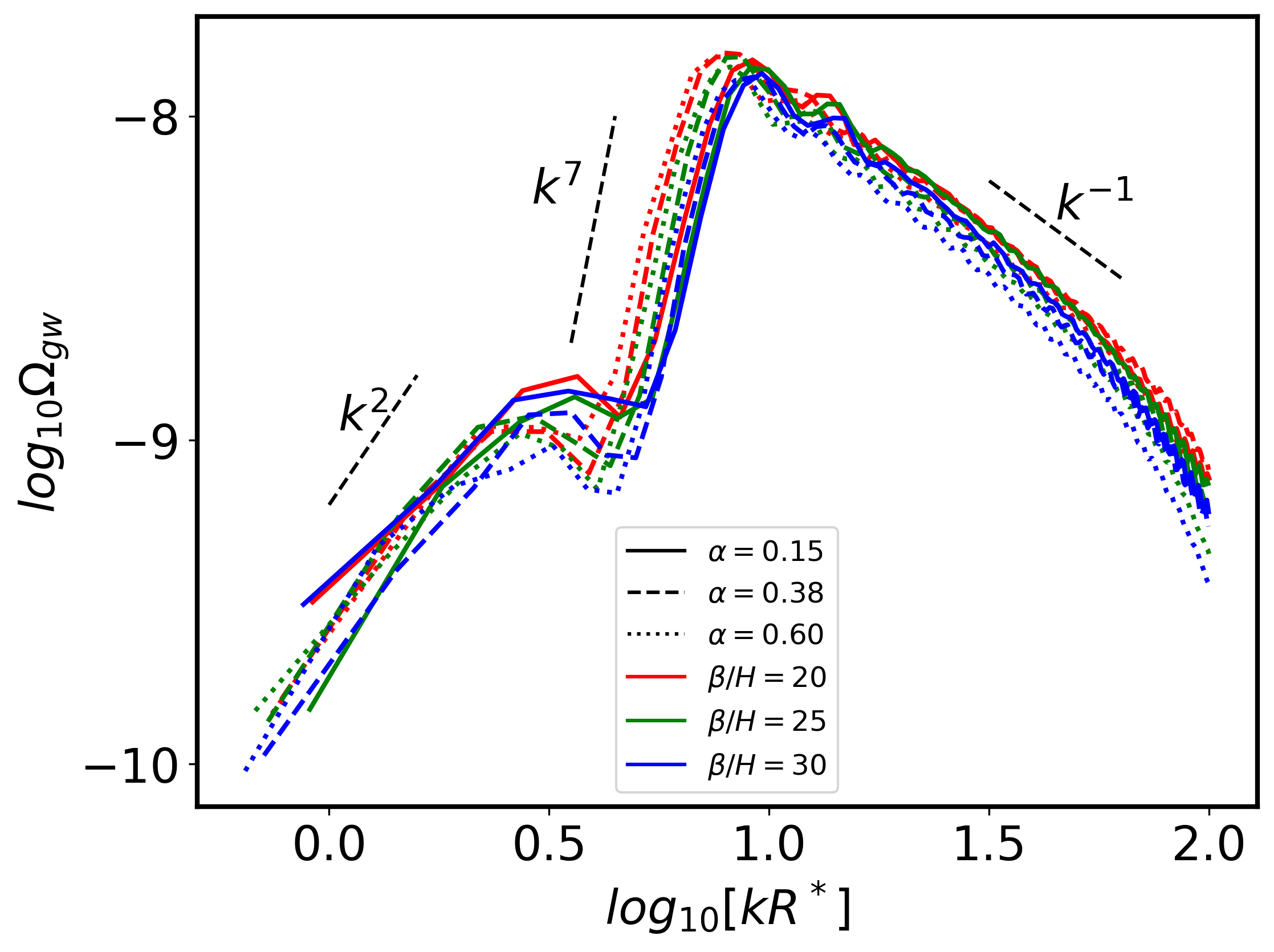}
   	\includegraphics[width=0.32\textwidth,clip]{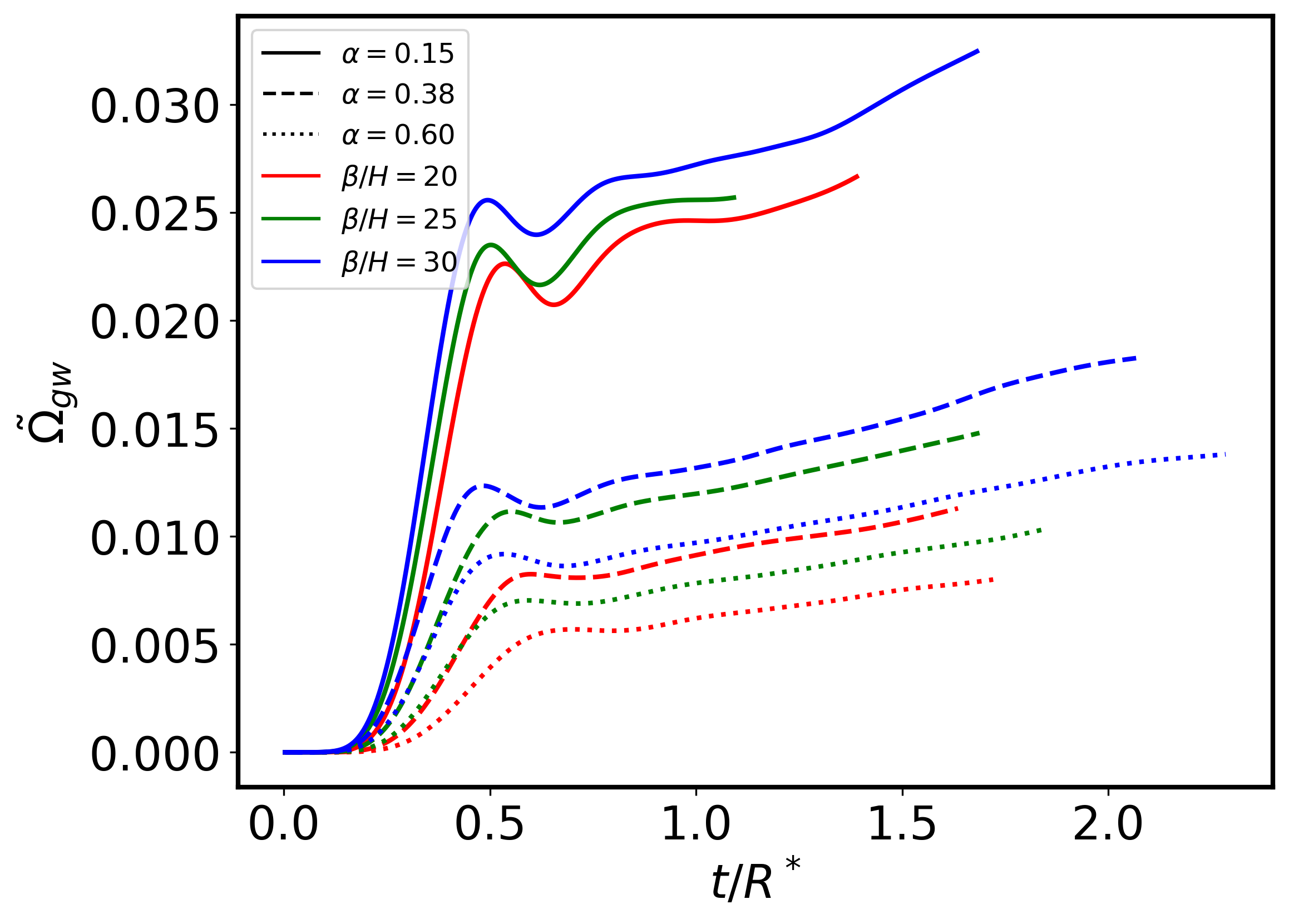}\\
    \includegraphics[width=0.32\textwidth]{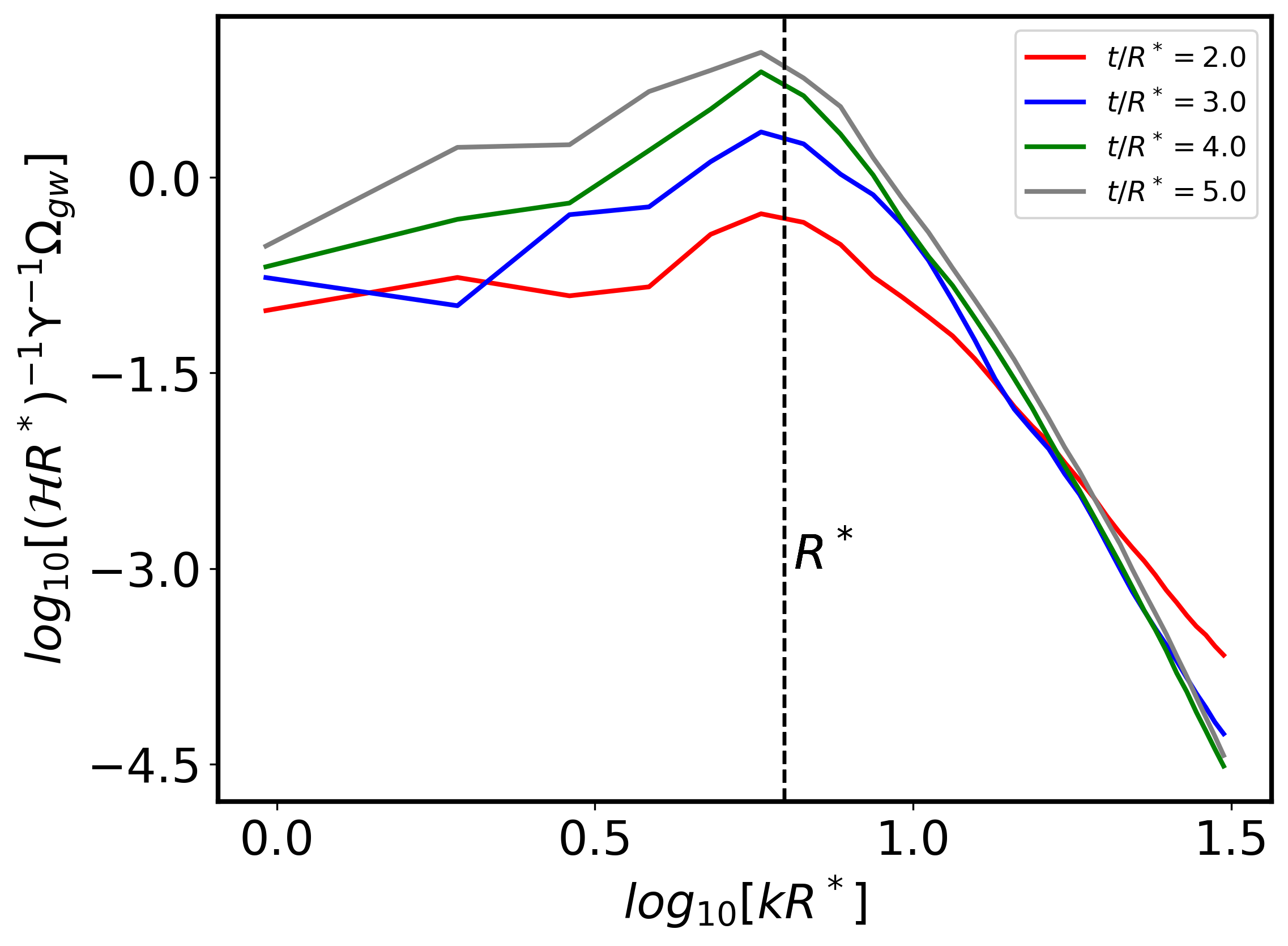}
    \includegraphics[width=0.32\textwidth,clip]{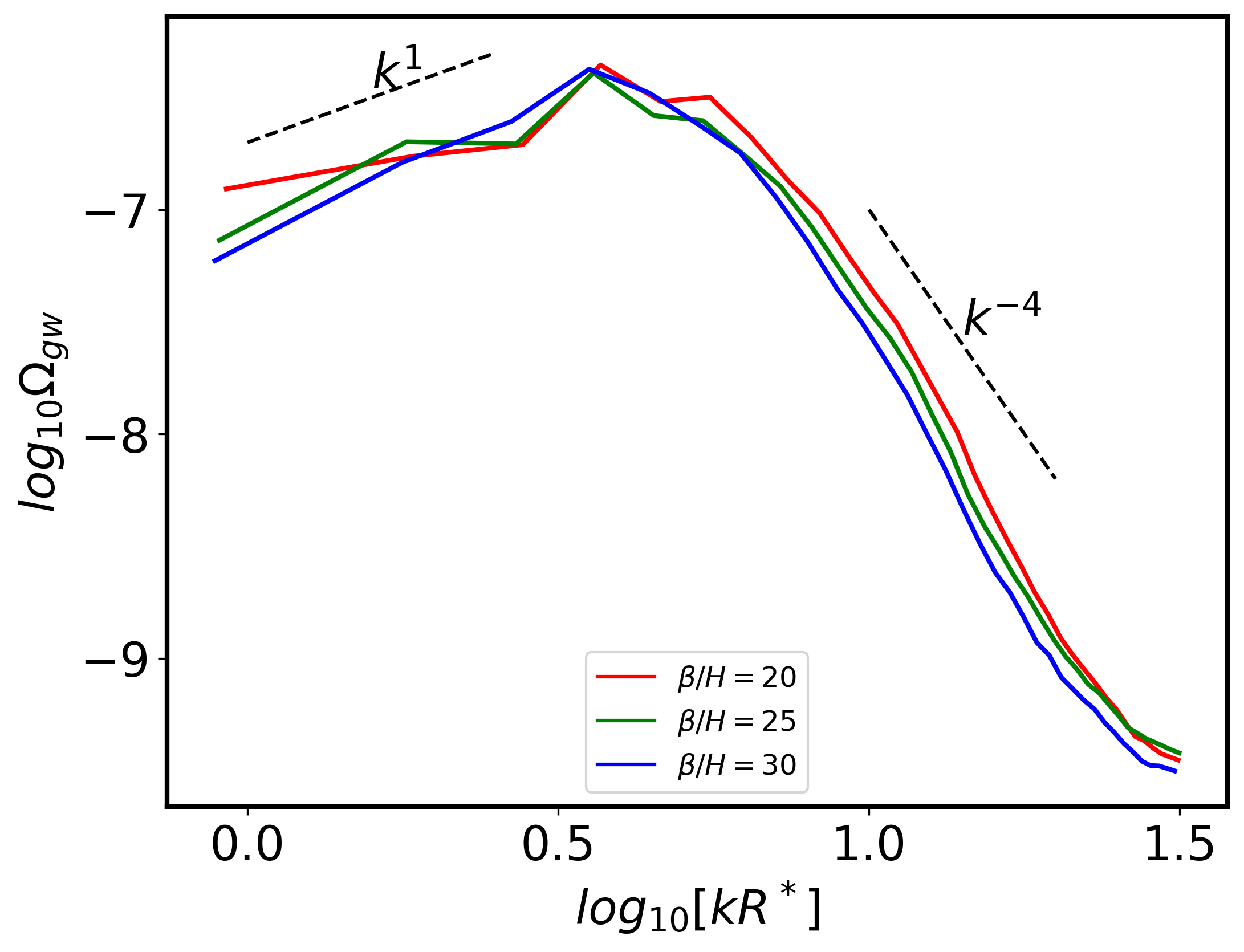}
    \includegraphics[width=0.32\textwidth,clip]{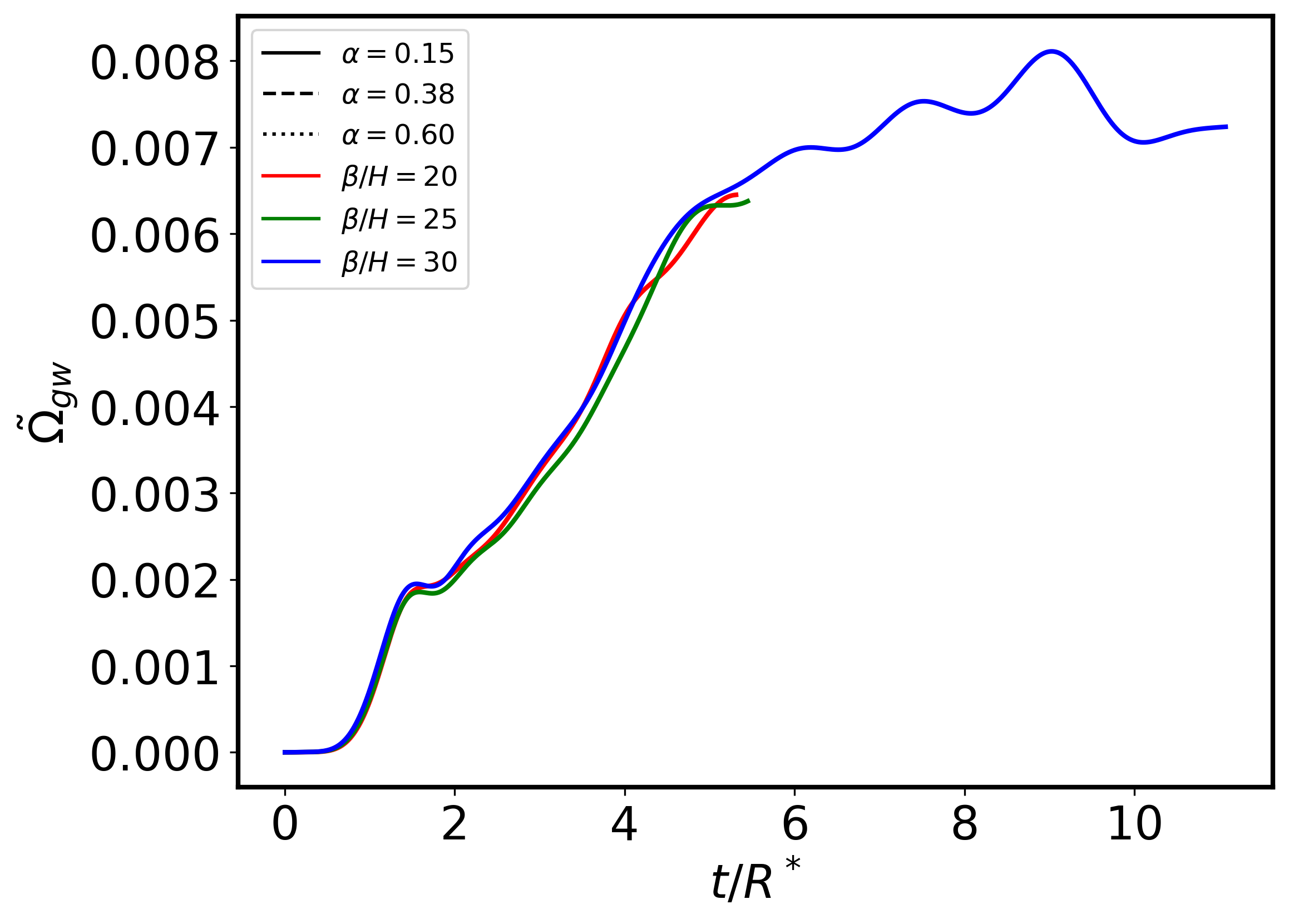}
    \caption{The normalized sound wave GW spectrum $(\mathcal{H}_n R^*)^{-1}\Upsilon^{-1}\Omega_{gw}$ for $\alpha_0=0.15$, $\beta/H=20$, the bubble-collision (field) GW spectra at the end of the phase transition, and the time evolution of the reduced sound wave spectra $\tilde{\Omega}_{gw}^f$ in the expanding Universe, for detonations (top row) and deflagrations (bottom row).}
\label{fig:GWs}
\end{figure*}

Fig.~\ref{fig:GWs} displays the gravitational-wave spectra associated with the phase transition, with the top row corresponding to detonations and the bottom row to deflagrations. The left panels show the sound wave (fluid) contribution to the GW spectrum for $\alpha_0 = 0.15$ and $\beta/H = 20$, with time slices at $t/R^* = 0.5$, $0.7$, $0.9$ and $1.0$ for the detonation and at $t/R^* = 2$, $3$, $4$, and $5$ for the deflagration. The normalized spectrum exhibits two peaks located at the two characteristic scales of the sound shell model, namely the sound-shell width $\Delta R^*$ and the mean bubble separation $R^*$. Both scales are marked by vertical dashed lines in the detonation panel, while only $R^*$ is marked in the deflagration panel since $\Delta R^*$ is close to $R^*$ there. The middle panels show the bubble-collision (field) contribution to the GW spectrum at the end of the phase transition, when $\omega \simeq 1/3$. The peak positions are the same as in the sound wave (fluid) contribution. In the ultraviolet region the detonation spectrum falls off as $k^{-1}$, consistent with the envelope approximation~\cite{Hindmarsh:2015qta}, while the deflagration spectrum falls off as $k^{-4}$. In all our simulations the sound wave contribution dominates over the bubble-collision (field) one. The right panels show the time evolution of the reduced sound wave spectrum $\tilde\Omega^f_{\rm gw}$, which approaches its asymptotic value as the phase transition proceeds, reaching $\tilde\Omega^f_{\rm gw}\simeq 10^{-2}$ at the end of the phase transition ($\omega\simeq 1/3$).
\begin{table*}[!htbp]
    \centering
    \renewcommand{\arraystretch}{1.2}
    \setlength{\tabcolsep}{5pt}  
    \caption{Summary of the simulation results for the detonation (upper rows) and deflagration (lower rows) runs. All quantities are evaluated at the end of the PTs ($\omega \simeq 1/3$).}
    \begin{tabular}{cc|cccccccccc}
        \hline
        $\alpha_0$  &  $ \beta/H$   &  $v_w$   &  $\eta$       &       $\kappa_{sw}$      &  $c_s$  & $|\nabla \times v|T_0^{-1}$  & $\overline{U}_f$ & $\overline{U}_\phi$ &  $\Upsilon$ & $(\mathcal{H}_n R^*)^{-1}(\Upsilon)^{-1}\Omega_{gw}^{f}$ & $\Omega_{gw}^{\phi}$ \\
        \hline
        $0.15$ & $20$ & $0.81$ & $0.248$ & $0.126$ & $0.534$ & $4.47\times10^{-5}$ & $0.093$ & $0.00314$ & $0.197$ & $4.11\times10^{-2}$ & $1.70\times10^{-8}$ \\
        $0.38$ & $20$ & $0.91$ & $0.294$ & $0.252$ & $0.516$ & $1.92\times10^{-4}$ & $0.209$ & $0.00286$ & $0.202$ & $4.48\times10^{-2}$ & $1.85\times10^{-8}$ \\
        $0.60$ & $20$ & $0.95$ & $0.315$ & $0.329$ & $0.505$ & $2.27\times10^{-4}$ & $0.287$ & $0.00248$ & $0.208$ & $4.44\times10^{-2}$ & $1.80\times10^{-8}$ \\
        $0.15$ & $25$ & $0.81$ & $0.244$ & $0.110$ & $0.536$ & $4.05\times10^{-5}$ & $0.087$ & $0.00311$ & $0.187$ & $4.05\times10^{-2}$ & $1.58\times10^{-8}$ \\
        $0.38$ & $25$ & $0.91$ & $0.285$ & $0.190$ & $0.520$ & $1.70\times10^{-4}$ & $0.181$ & $0.00240$ & $0.178$ & $5.67\times10^{-2}$ & $1.64\times10^{-8}$ \\
        $0.60$ & $25$ & $0.95$ & $0.306$ & $0.243$ & $0.510$ & $1.72\times10^{-4}$ & $0.246$ & $0.00256$ & $0.183$ & $5.39\times10^{-2}$ & $1.55\times10^{-8}$ \\
        $0.15$ & $30$ & $0.81$ & $0.242$ & $0.099$ & $0.536$ & $1.01\times10^{-4}$ & $0.082$ & $0.00291$ & $0.179$ & $4.48\times10^{-2}$ & $1.48\times10^{-8}$ \\
        $0.38$ & $30$ & $0.91$ & $0.280$ & $0.159$ & $0.521$ & $1.39\times10^{-4}$ & $0.166$ & $0.00225$ & $0.163$ & $5.61\times10^{-2}$ & $1.42\times10^{-8}$ \\
        $0.60$ & $30$ & $0.95$ & $0.299$ & $0.193$ & $0.513$ & $1.49\times10^{-4}$ & $0.220$ & $0.00221$ & $0.164$ & $6.50\times10^{-2}$ & $1.37\times10^{-8}$ \\
        \hline
        $0.15$ & $20$ & $0.27$ & $6.192$ & $0.598$ & $0.458$ & $8.56\times10^{-5}$ & $0.202$ & $0.00352$ & $0.511$ & $4.59\times10^{-2}$ & $5.09\times10^{-7}$ \\
        $0.38$ & $25$ & $0.27$ & $6.198$ & $0.563$ & $0.458$ & $8.45\times10^{-5}$ & $0.196$ & $0.00313$ & $0.512$ & $4.62\times10^{-2}$ & $4.68\times10^{-7}$ \\
        $0.60$ & $30$ & $0.27$ & $6.201$ & $0.519$ & $0.459$ & $8.35\times10^{-5}$ & $0.188$ & $0.00304$ & $0.512$ & $4.96\times10^{-2}$ & $4.72\times10^{-7}$ \\
        \hline
    \end{tabular}
    \label{tab:summary}
\end{table*}

We list some of the simulation parameters in Tab.~\ref{tab:summary}. For each parameter set ($\alpha_0$, $\beta/H$) at the initial temperature $T_0$, we list the bubble wall velocity $v_w$, the friction coupling $\eta$, the sound wave efficiency factor $\kappa_{sw}$, the sound speed $c_s$, the averaged magnitude of the velocity curl $|\nabla \times v|T_0^{-1}$, the root-mean-square velocities $\overline{U_f}$ and $\overline{U_\phi}$ of the fluid and the field, the growth factor $\Upsilon$, and the resulting GW amplitudes.

\end{document}